\documentclass[10pt,journal,compsoc]{IEEEtran}
\makeatletter
\def\endthebibliography{%
  \def\@noitemerr{\@latex@warning{Empty `thebibliography' environment}}%
  \endlist
}
\makeatother

\usepackage{setspace}
\ifCLASSOPTIONcompsoc
  \usepackage[nocompress]{cite}
\else
  \usepackage{cite}
\fi
\ifCLASSINFOpdf
   \usepackage[pdftex]{graphicx}
\else
\fi
\usepackage{url}
\usepackage{hyperref}
\usepackage{float}
\usepackage[dvipsnames]{xcolor}
\usepackage{wrapfig}
\begin{document}

\title{Survey on Visual Analysis of \\ Event Sequence Data}
\author{Yi Guo, Shunan Guo, Zhuochen Jin, Smiti Kaul, David Gotz, and~Nan Cao% <-this % stops a space
\IEEEcompsocitemizethanks{

\IEEEcompsocthanksitem Yi Guo,  Shunan Guo, Zhuochen Jin, and Nan Cao are with the Intelligent Big Data Visualization Lab, Tongji University, Shanghai, China. Nan Cao is the corresponding author.
E-mail: nan.cao@gmail.com

\IEEEcompsocthanksitem Smiti Kaul and David Gotz are with the Visual Analysis and Communication Lab at the University of North Carolina at Chapel Hill, NC, USA. E-mail: {smiti, gotz}@unc.edu

}}

\IEEEtitleabstractindextext{%

\begin{abstract}
Event sequence data record series of discrete events in the time order of occurrence. They are commonly observed in a variety of applications ranging from electronic health records to network logs, with the characteristics of large-scale, high-dimensional and heterogeneous. This high complexity of event sequence data makes it difficult for analysts to manually explore and find patterns, resulting in ever-increasing needs for computational and perceptual aids from visual analytics techniques to extract and communicate insights from event sequence datasets. In this paper, we review the state-of-the-art visual analytics approaches, characterize them with our proposed design space, and categorize them based on analytical tasks and applications. From our review of relevant literature, we have also identified several remaining research challenges and future research opportunities.
\end{abstract}
}

\maketitle

% is not selected <OR> if conference mode is selected - because compsoc
% conference papers position the abstract like regular (non-compsoc)
\IEEEdisplaynontitleabstractindextext

% \fi
\IEEEpeerreviewmaketitle

\ifCLASSOPTIONcompsoc
\IEEEraisesectionheading{\section{Introduction}\label{sec:introduction}}
\else
\section{Introduction}
\label{sec:intro}
\fi

\IEEEPARstart{E}{vent} sequence data are found across a vast array of applications and domains.  In fields as diverse as computer security, advertising, and healthcare, discrete observations of different types are collected over time and arranged in sequence based on the specific entity for which the event is germane. For example, network logs in computer systems capture timestamped sequences of events (logins, requests, faults, etc.) for specific devices. Similarly, clickstreams used to tailor advertising capture sequences of interaction events for individual users as they navigate websites. Electronic health records, meanwhile, capture events (e.g., diagnoses, procedures) over time for individual patients.  The ubiquity of event sequence data reflects both (1) the relative ease with which it can be captured, and (2) the desire to leverage this form of data to gain new insights about real-world systems.

These common goals, however, are challenged by the great heterogeneity that exists within different properties of event sequence data and the types of insights that are sought.  For example, event sequences can be high-dimensional (with many event types) or low-dimensional (very few types of events).  They can be sparse and irregular over time, or dense and evenly spaced.  Events can have zero attributes or many, can be point events or intervals, and can be strictly sequential or occur in parallel.  Similarly, the types of analysis tasks can vary widely based on the types of insights one seeks.  Are analysts interested in common patterns or rare outliers? Are analysts focused on prediction, or identification of predictive factors for intervention?  Are analysts examining a single sequence or comparing across multiple sets of sequences in aggregate?  These are just a few examples of the wide variety of data and task challenges which present themselves in 
event sequence analysis.

These difficult and diverse methodological challenges have motivated a broad range of recent research activities which aim to solve one or more aspects of the event sequence analysis problem. This has led, in turn, to a proliferation of different visual analysis methods and prototypes, each of which has distinct capabilities and advantages in certain contexts.  This has resulted in a situation where the state-of-the-art for event sequence data is often difficult to discern.  The latest research often offers multiple visual analytics approaches for specific types of challenges.  Moreover, the same solution may be effective at addressing difficulties that stem from two or more different challenges. Yet in other cases, open problems remain unaddressed.

The aim of this survey is to provide a comprehensive review and characterization of the state-of-the-art in visual analytics research for event sequence data.  Through the collection and analysis of the literature on this topic, we identify key dimensions of the event sequence visual analytics design space. We then use those dimensions, as well as a characterization of different types of event sequence analysis tasks, to organize existing methods and identify common approaches to specific targeted problems.  Moreover, we identify areas with little prior work which remain a challenge for future research.

This literature review represents the first (to our knowledge) comprehensive attempt to survey and characterize event sequence data visual analytics methods.  In this way, this review promises to help researchers understand key dimensions that unify prior work, how prior research fits together within this complex design space, and which event sequence data analysis challenges remain insufficiently addressed.  Moreover, the results can provide value to practitioners as an organized catalog of alternative approaches that are most appropriate for specific types of event sequence data problems. We developed a web-based survey browser \footnote{\href{http://eventvis.idvxlab.com/}{http://eventvis.idvxlab.com/}} to facilitate the exploration of our created taxonomy and reviewed techniques.

\section{Related Surveys and Methodology}
\label{sec:method}
In this section, we first discuss survey papers that are relevant to this work, and then introduce our methodology of selecting papers and creating our taxonomy.
\vspace{-3.5mm}
\subsection{Related Surveys}
This section provides an overview of the surveys that are relevant to visual analysis of event sequence data. Keim et al.~\cite{keim2008visual} proposed a definition and an analytical pipeline for visual analytics, which inspires our formalization of the design space that we discuss later in Section~\label{sec:design}. A prior survey by Sun et al.~\cite{sun2013survey} generalized visual analytics techniques by different data types, among which the review of visual analytics approaches for temporal data is most relevant to our work. Our work, by contrast, focus on a more specific type of temporal data -- event sequence data. In addition, some scholars attempted to dive into particular visual forms or visual analytics approaches for a single analytical task that are partially related to our survey. For example, Brehmer et al. \cite{brehmer2016timelines} formalized the design space for a representative form for visualization of event sequence data -- timeline-based visualizations. Jentner and Keim~\cite{jentner2019visualization} reviewed visualization and visual analytics methods for exploring frequent patterns. Given the broad application of event sequence data, we also notice a larger group of surveys linked to applications where event sequences are commonly collected, such as social media data~\cite{wu2016survey}, smart manufacturing ~\cite{zhou2019survey}, and anomalous user behaviors ~\cite{shi2019visual}. Different from existing work that summarizes techniques for a particular visualization, visual analytical task, or application related to temporal event sequence, our work aims to provide a more holistic overview of the visual analytics approaches for all types of event sequence data so as to benefit practitioners from a wider range of applications.

\vspace{-4mm}
\subsection{Survey Methodology}
\label{sec:methodology}
This survey aims to obtain an overview of existing visual analytics techniques that are developed for event sequence data. To construct a structured and comprehensive taxonomy, we start by formalizing a design space for developing visual event sequence analysis tools (discussed in Section~\ref{sec:design_space}). In particular, we leverage the conventional visual analytics pipeline~\cite{keim2008visual} followed by most visual analytics techniques, which revolves around four key components: data, model, visualization, and knowledge. Since deriving knowledge from model and visualizations can be subjective and difficult to standardize, we exclude knowledge inference from the scope of our design space. In addition, user interaction that links the components throughout the pipeline is also indispensable in the visual analytics process. These considerations led to our final proposed design space with with the following four dimensions: \emph{data scales}, \emph{analysis techniques}, \emph{visual representations}, and \emph{interactions}. For each dimension, we further enumerate all alternatives as we conduct our review of the existing studies.

We collect relevant papers from visualization journals and conferences. We followed two main approaches when collecting the papers: reference-driven and search-driven selections. For the reference-driven selection, we utilized a core set of state-of-the-art techniques in this topic known to us in advance as a starting point, and extended the range of work by going through cited and citing publications. 

For the search-driven selection, we went through two rounds of paper collection. The first round involves a coarse search of event sequence analysis and visualization techniques from high impact conferences and journals in the field of information visualization and data mining. In particular, we select six visualization conferences: IEEE VAST, IEEE InfoVis, ACM CHI, ACM IUI, EG/IEEE EuroVis and IEEE PacificVis, three visualization journals: IEEE TVCG, IEEE CG\&A, Computer Graphics Forum, four data mining conferences: NeruIPS, WWW, ACM SIGKDD, ICML and two journals: IEEE TKDD, ACM TIST. We used two search queries: "event sequence" \verb AND ~"analysis"; "event sequence" \verb AND ~"visualization" to collect papers broadly, then reviewed the abstracts and full texts to finalize our selection. We labeled each work with their correspondence in each dimension respectively. Note that event sequence analysis techniques are only labeled in the first two dimensions. In addition, according to Keim et al.~\cite{keim2008visual}, the choices of analysis methods, visual representations, and interactions depend on the analytical tasks and application scenarios. Therefore, we also label each collected publication with their motivated tasks and applications. This gives us a full list of nine analytical tasks, which we further organized in to five categories as outlined in Section~\ref{sec:vis}, and seven applications under three major categories as outlined in Section~\ref{sec:app}. 

% This resulted in a total of 153 most relevant publications of event sequence analysis, 89 publications of event sequence visualization and visual event sequence analysis. 
For each analytical task and application, we went through another round of complementary paper collection for visualization and visual analytics techniques with search queries that combines specific tasks or applications, such as "event sequence summarization" \verb AND  "visualization", "medical data" \verb AND  "visualization", etc. The entire selection process ended up with 153 most relevant publications of event sequence analysis, and 144 publications of event sequence visualization and visual event sequence analysis. We further refined our selection with 100 most representative and up-to-date event sequence visualization and visual analytics studies to discuss in this paper. Additionally, this survey also includes the review of 8 related surveys, 5 event sequence analysis techniques, 9 visualization techniques in the field of causality analysis yet are not related to event sequence data, and referred to 10 papers regarding the theory of visual analytics, research challenges and oppotunities of visual event sequence analysis. This result in a total of 133 papers that are covered in this survey.

\indent 
The remaining survey is organized as follows. We first introduce the taxonomy of our survey by formalizing our proposed design space and outline visual analytical tasks in Section \ref{sec:design}. Section~\ref{sec:vis} elaborates the state-of-the-art solutions developed for each analytical task respectively through an analysis of their corresponding design components of the design space. Then, we provide an overview of applications where event sequence data are commonly observed in Section~\ref{sec:app}, serving as a more direct guide to practitioners of visual analytics techniques. Finally, we discuss our reflections on research challenges, opportunities in Section \ref{sec:discussion} and conclude our work in Section \ref{sec:conclusion}.

\vspace{-3.5mm}
\section{Taxonomy}
\label{sec:taxonomy}
In this section, we introduce the design space and the collection of visual analysis tasks built from the processes of paper gathering and labeling as mentioned in Section ~\ref{sec:methodology}. The design space and visual analytical tasks form a taxonomy that we further use to structure the survey. Specifically, in Section~\ref{sec:vis}, we partition visual analytics techniques based on their primary analytical task under a consideration that most visual analytics systems are developed around a single analytical task. For each analytical task, we further characterize the relevant papers by exposing the dimensions in the design space that are leveraged to develop each visual analytics method. We also discuss the applications of the visual event sequence analysis techniques in Section~\ref{sec:app} to provide domain-specific guidance for practitioners. The applications, however, are not included in our taxonomy, because most of the techniques we collected are developed for event sequence analysis in general cases rather than a specific application.

\vspace{-3mm}
\subsection{Design Space}
\label{sec:design_space}
In the following, we introduce the dimensions of our design space and highlight the key elements (i.e., data scales or techniques) in each dimension that are frequently used for designing and building a visual analysis system for analyzing event sequence data. 
\begin{figure*}[!ht]
	\centering    
	\includegraphics[width=\linewidth]{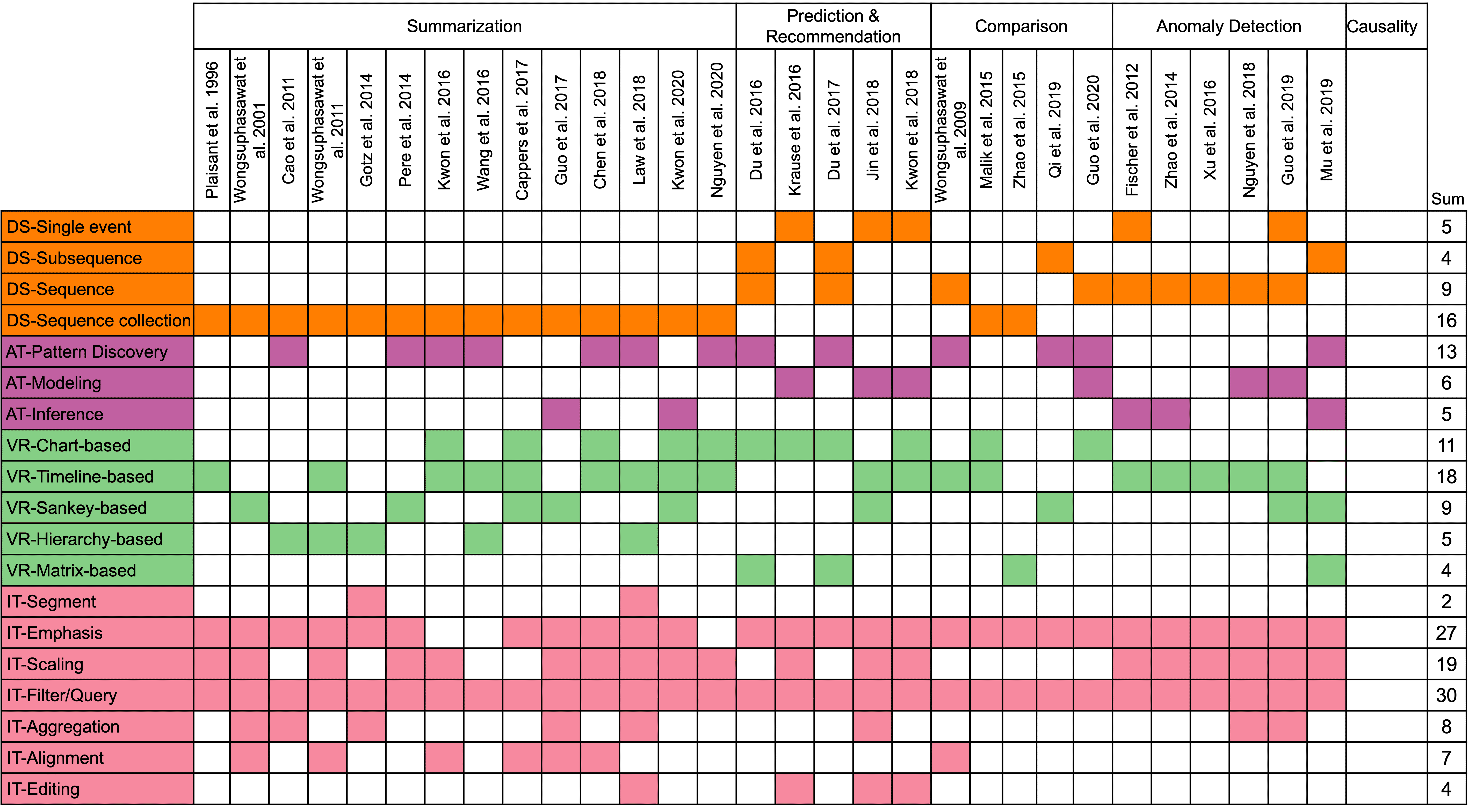}
	\vspace{-8mm}
	\caption{
	The most cited papers regarding event sequence visualization and visual analytics techniques grouped by different tasks. Each paper is labeled by the relevant design elements in the design space. The rows are grouped and colored by dimensions of our proposed design space: DSs - Data Scales; ATs - Analysis Techniques; VRs - Visualization Representations; ITs - Interaction Techniques.
	} 
	\vspace{-3mm}
	\label{figure:table1}
\end{figure*}
\begin{figure}[!ht]
	\centering    
	\includegraphics[width=\linewidth]{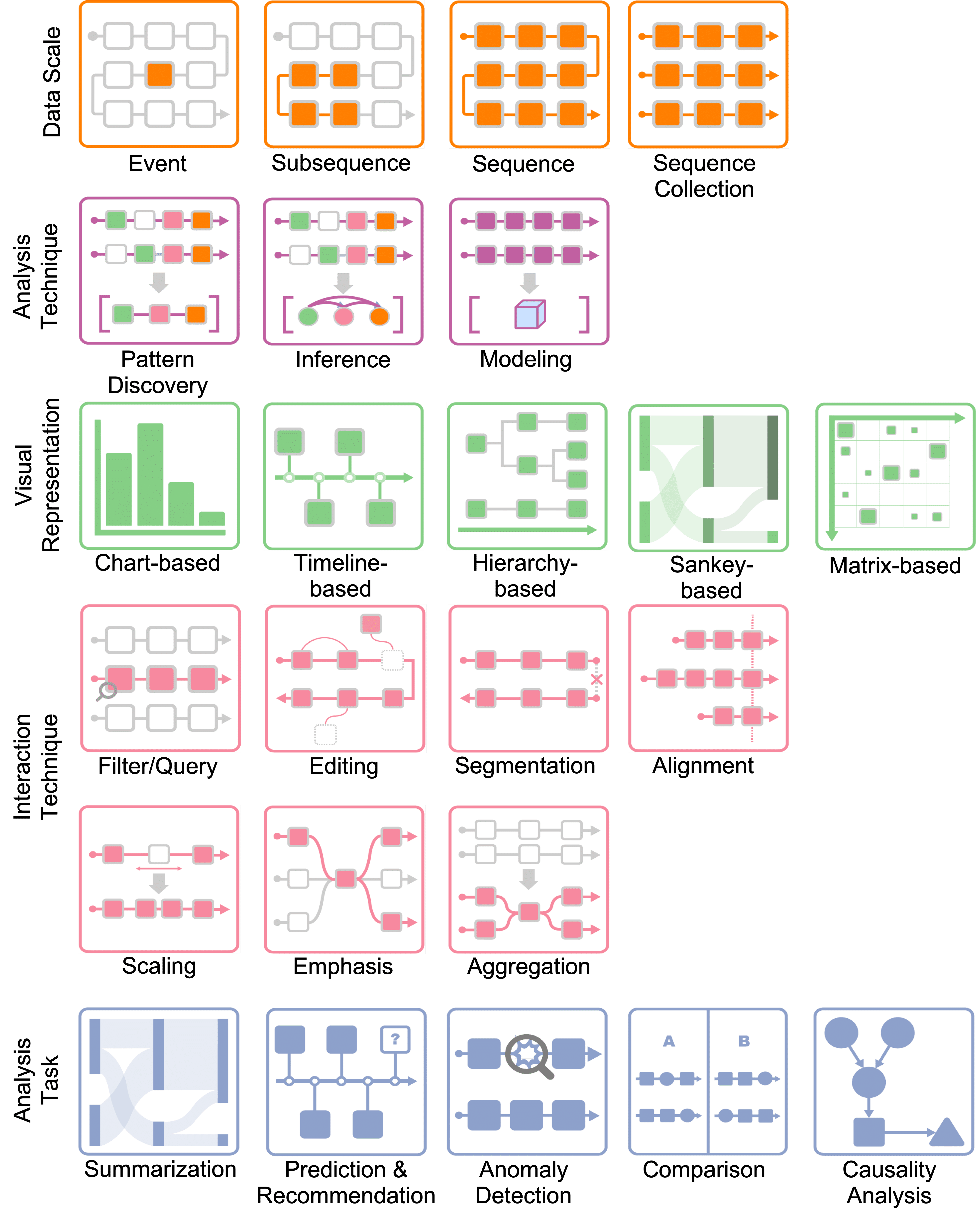}
	\vspace{-7mm}
	\caption{The taxonomy of this survey is established using a design space for developing visual analysis methods for event sequences with four dimensions: data scale, analysis technique, visual representation, interaction technique, and a organized catelog of visual analytics tasks that includes summarization, prediction \& recommendation, anomaly detection, comparison, causal analysis.}
	\vspace{-5mm}
	\label{figure:space}
\end{figure}
\vspace{-2mm}
\subsubsection{Dimension 1: Data Scales}
Our proposed design space starts with identifying the granularity of data that the visual analysis is able to cover. For a given event sequence dataset, we summarize the following levels of data granularity.

\begin{wrapfigure}[3]{l}{0.042\textwidth}
\centering
\includegraphics[width=0.059\textwidth]{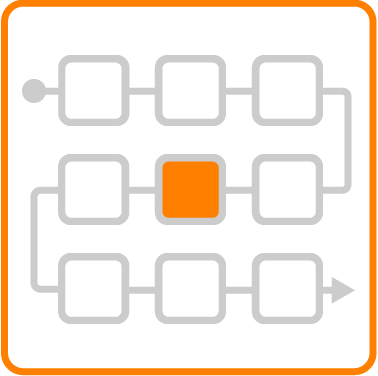}
\end{wrapfigure}
\noindent \textbf{Event:} Individual events represent the finest granularity of event sequences. Each event can be characterized by attributes such as event type, time of occurrence, and duration. Visual analytics techniques often attempt to drill down to individual events to provide users with low-level details of the analysis result. For example, Vistracker~\cite{fischer2012vistracer} identifies anomalous events in trace routes based on event attributes. Carepre~\cite{carepre} predicts upcoming disease based on historical sequence of medical events. 

\begin{wrapfigure}[3]{l}{0.042\textwidth}
\centering
\includegraphics[width=0.059\textwidth]{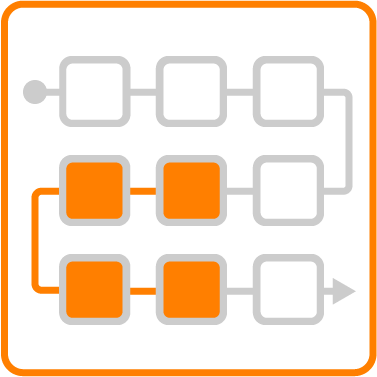}
\end{wrapfigure}
\noindent \textbf{Subsequence:} Subsequences are segments of event sequences with the temporal order of events being preserved. Meaningful subsequences can represent the major characteristic of the sequence. EventAction~\cite{du2016eventaction} utilizes the number of common subsequences between individuals to measure sequence similarities. MOOCad.~\cite{mu2019moocad} leverage anomalous frequent subsequences to facilitate the reasoning of sequence anomalies. 

\begin{wrapfigure}[3]{l}{0.042\textwidth}
\centering
\includegraphics[width=0.059\textwidth]{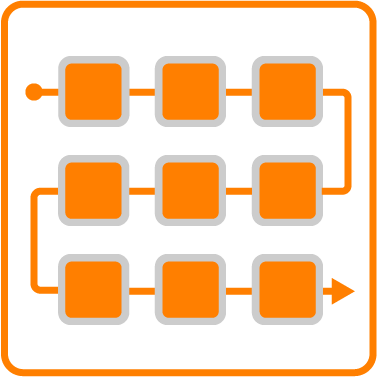}
\end{wrapfigure}
\noindent \textbf{Sequence:} An event sequence is the complete record of events that are performed or experienced by a sequence entity (e.g., a patient or a customer). The entire sequence is often analyzed when attempting to get a complete view of the entity's experience.  In~\cite{zhao2014fluxflow,guo2019visual,nguyen2018understanding}, anomalous entities are detected by analyzing their corresponding progressions of events. Similarly, Guo et al.~\cite{guo2020comparative} utilize the embedding of each sequence to estimate the similarity between entities.

\begin{wrapfigure}[3]{l}{0.042\textwidth}
\centering
\includegraphics[width=0.059\textwidth]{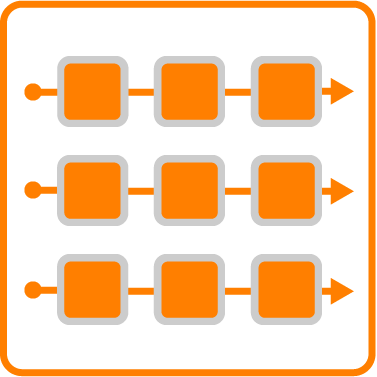}
\end{wrapfigure}
\noindent \textbf{Sequence Collection:} A collection of sequences are analyzed when summarizing common patterns in the dataset or comparing different groups of sequences. For example, visual sumarization techniques \cite{plaisant1996lifelines, perer2014frequence,guo2017eventthread,guo2018visual} aim to provide a summary of patterns and identify entities with common progressions in a collection of sequences. MatrixWave~\cite{zhao2015matrixwave} is designed to compare two collections of event sequences and analyze their differences.

\vspace{-1mm}
\subsubsection{Dimension 2: Analysis Techniques}
Visual analytics techniques for event sequence data are incorporated with back-end data mining algorithms to support complex analytical tasks. Based on a review of event sequence analysis methods, we identify the following analysis techniques.

\begin{wrapfigure}[3]{l}{0.042\textwidth}
\centering
\includegraphics[width=0.059\textwidth]{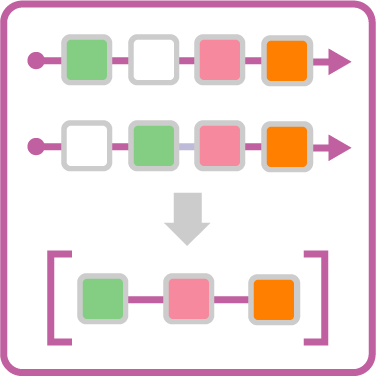}
\end{wrapfigure}
\noindent \textbf{Pattern discovery:} Pattern discovery aims to find frequently occurring patterns and statistically significant associations of data samples. In the analysis of event sequence data, pattern discovery techniques can be further categorized into frequent pattern mining techniques and similarity analysis techniques based on different analytical goals. Frequent pattern mining techniques are used to uncover common subsequences in the event sequence dataset. For instance, Perer et al.~\cite{perer2014frequence} proposed a visual analytics system that employs a SPAM-based algorithm to extract frequent patterns in a collection of event sequences. Similarity analysis techniques utilize event patterns of each sequence to quantify the similarity between sequences. For example, in Eventaction~\cite{du2016eventaction} and Similan~\cite{wongsuphasawat2009finding}, two different similarity measurements were proposed based on 
commonness and differences between events across different event sequences.
\begin{wrapfigure}[3]{l}{0.042\textwidth}
\centering
\includegraphics[width=0.059\textwidth]{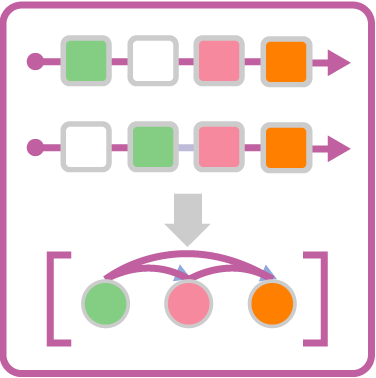}
\end{wrapfigure}

\noindent \textbf{Sequence inference:} Inference is the process of drawing conclusions based on evidence observed in existing data. Conclusions derived from inference techniques are tenable under certain conditions but can be incorrect when applied to unobserved data. Existing inference techniques for event sequence analysis mainly include self-exciting point process and graphical model.
Self-exciting point process is a probabilistic model that describes the occurrence probabilities of events over time. The occurrence of upcoming events is influenced by historical events. For example, Hawkes Process is widely employed to model sequential data under the assumption that the impact of the previous event is approximated by a numerical integration over time~\cite{xu2016learning,mei2017neural}.
Graphical model, on the other hand, presents the conditional dependence between events with a event correlation graph, such as Bayesian Networks~\cite{bhattacharjya2020event} and Markov Chain~\cite{stopar2018streamstory}.

\begin{wrapfigure}[3]{l}{0.042\textwidth}
\centering
\includegraphics[width=0.059\textwidth]{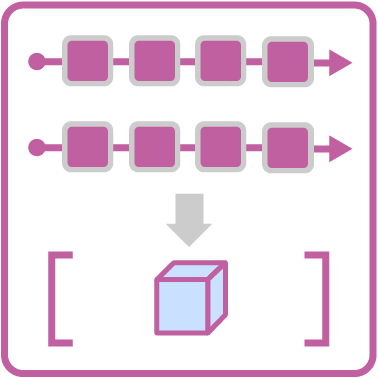}
\end{wrapfigure}
\noindent \textbf{Sequence modeling:} While sequence inference techniques are not capable of making predictions on unobserved data, sequence modeling methods are developed to build a reliable model to characterize observed data while ensuring the model's generalization abilities on unobserved data. Event sequence models are generally specifically designed depending on the analytical tasks, such as classification (e.g., support vector machines, decision trees) and clustering (e.g, k-means). Neural network models, especially recurrent neural networks (RNN) are also commonly applied to model event sequences due to their inherent sequential structure and superior performance comparing to traditional machine learning model. For instance, CarePre~\cite{carepre} employed attention-based RNNs to predict upcoming events based on historical events in sequences, and Guo et al.~\cite{guo2019visual} embedded RNNs into Variational Auto-Encoder to detect anomalous sequences in the dataset.

% \begin{itemize}
%     \item {Pattern Discovery}
%     \item {Modeling}
%     \item {Inference}
% \end{itemize}

\vspace{-3mm}
\subsubsection{Dimension 3: Visual Representations}
Existing visual analytics techniques leverage a variety of visual representations to display event sequence data and communicate insightful patterns. The visual representations also determine how events and sequences are organized and aggregated. We identify five categories of visual representation for displaying event sequence data as follows. 

\begin{wrapfigure}[3]{l}{0.042\textwidth}
\centering
\includegraphics[width=0.059\textwidth]{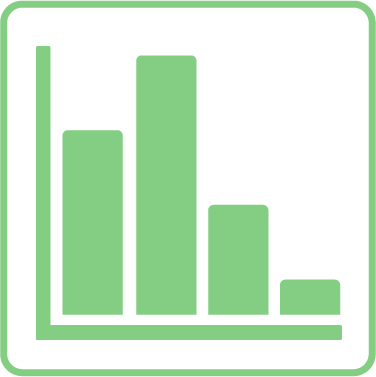}
\end{wrapfigure}
\noindent\textbf{Chart-based visualizations:} Visualization charts, such as bar charts and scatter plots, are commonly used to display event features and event distributions of event sequences. For instance, Coco~\cite{malik2016high} uses a table to compare event distributions of two different groups of sequences and a scatter plot to show the number of records containing particular events or subsequences.

\begin{wrapfigure}[3]{l}{0.042\textwidth}
\centering
\includegraphics[width=0.059\textwidth]{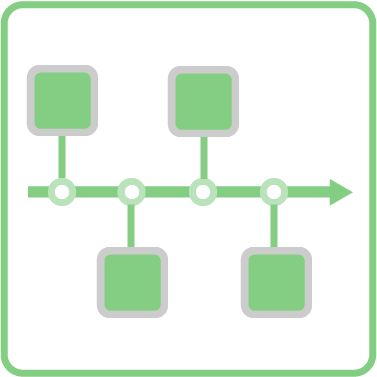}
\end{wrapfigure}
\noindent\textbf{Timeline-based visualizations:} Timelines are the most intuitive visualizations that organize events of individual sequences successively in a temporal order. Events are generally represented with icons encoded by color, size, or shape to distinguish events with different attributes. For example, VASABI~\cite{vasabi} visualizes a sequence as a row of squares colored by event categories. 

\begin{wrapfigure}[3]{l}{0.042\textwidth}
\centering
\includegraphics[width=0.059\textwidth]{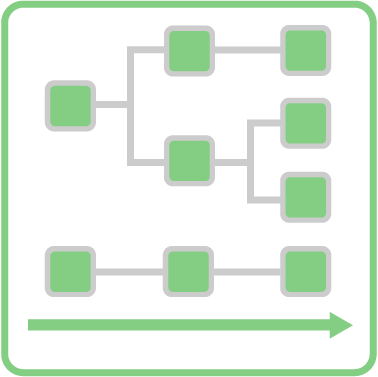}
\end{wrapfigure}
\noindent\textbf{Hierarchy-based visualizations:} In hierarchy-based visualizations, sequences are aggregated into a tree of sequences~\cite{wongsuphasawat2011lifeflow}, where each node represents an event placed according to its prefix in the sequence. A variety of visualization designs can be used to display this hierarchical structure of sequences, such as treemaps~\cite{trumper2012viewfusion}, node-link tree~\cite{vrotsou2009activitree}, and icicle plots \cite{liu2017coreflow,law2018maqui}. 

\begin{wrapfigure}[3]{l}{0.041\textwidth}
\centering
\includegraphics[width=0.059\textwidth]{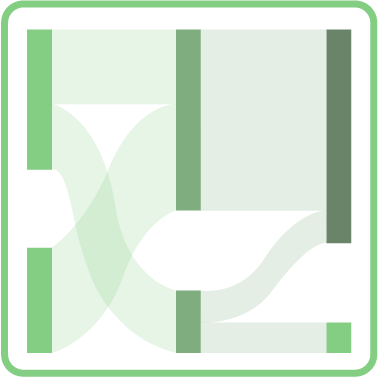}
\end{wrapfigure}
\noindent\textbf{Sankey-based visualizations: } The Sankey-based visualizations organize sequences into the structure of a Sankey diagram~\cite{riehmann2005interactive}. Instead of aggregating sequences into a tree structure as the hierarchy-based visualizations, Sankey-based methods aggregate sequences into a graph, focusing more on providing an overview of transitions between different types of events. Sankey-based visualizations can be further categorized into two different types of design. The first one is the directed node-link graph in which events are represented by nodes and transitions between events are represented by links~\cite{carepre,guo2018visual}. 
The second one is the traditional Sankey diagram, in which links are further encoded by width, representing the proportions of flow that split and merge among events~\cite{wongsuphasawat2011outflow,guo2017eventthread}. 

\begin{wrapfigure}[3]{l}{0.042\textwidth}
\centering
\includegraphics[width=0.059\textwidth]{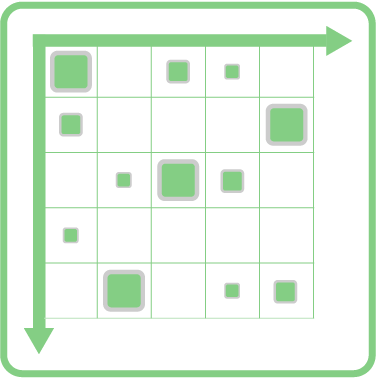}
\end{wrapfigure}
\noindent\textbf{Matrix-based visualizations:} Matrix-based visualizations are typically used to demonstrate a summary of event frequency or frequent patterns. For example, EventAction~\cite{du2016eventaction} incorporates an event matrix to summarize frequencies of events across different time intervals. Mu et al.~\cite{mu2019moocad} applied a matrix-based design to present lists of frequent activity patterns in each stage of sequence progression. In addition, a matrix-based design is also utilized to display frequencies of event transitions. For example, Zhao et al.~\cite{zhao2015matrixwave} transformed the traditional Sankey diagram into a sequence of matrices to display step-to-step transitions of web clickstream data.

\vspace{-2mm}
\subsubsection{Dimension 4: Interactions}
Visual analytics systems usually incorporate rich interactions to empower end users with sufficient flexibility and depth in data analysis. In the following, we summarize seven interaction techniques that are commonly applied in visual analytics systems for event sequence data.

\begin{wrapfigure}{l}{0.042\textwidth}
\centering
\includegraphics[width=0.059\textwidth]{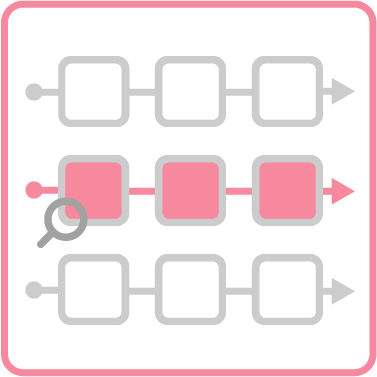}
\end{wrapfigure}
\noindent \textbf{Filter/query} allows users to make domain-specific data adjustment or selection based on certain conditions, so as to eliminate noisy and irrelevant data for better analytical performance. The types of filters include \emph{event filters} for filtering specific event types (e.g., \cite{guo2017eventthread,zhao2015matrixwave}), \emph{time filters} (e.g., \cite{law2018maqui,fischer2012vistracer}) for narrowing down to a range of time in the middle of the sequence for exploration, \emph{attribute filters} (e.g., \cite{chen2018sequence}) for retrieving a subset of event sequences based on sequence or event attributes, and \emph{pattern filters} (e.g., \cite{perer2014frequence,guo2019visual}) for querying event sequences that contain specific subsequences.

\begin{wrapfigure}{l}{0.042\textwidth}
\centering
\includegraphics[width=0.059\textwidth]{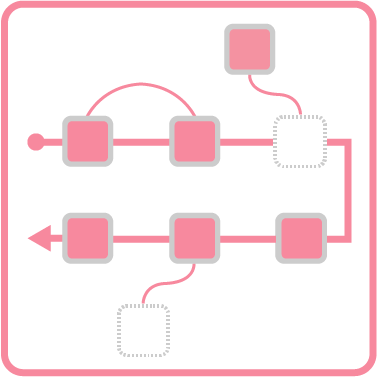}
\end{wrapfigure}
\noindent
\textbf{Editing} enables users to modify event sequences through 
\emph{adding new events}, \emph{removing existing events}, \emph{editing event order}, and \emph{editing event duration}, which is commonly employed in what-if simulation of event sequence predictions. The goal is to interactively explore the influence of historical events on the prediction results. For instance, in the CarePre \cite{carepre} and RetainVis \cite{kwon2018retainvis}, users can edit event sequences to understand how the change of individual events affects the prediction of risks. 

\begin{wrapfigure}{l}{0.042\textwidth}
\centering
\includegraphics[width=0.059\textwidth]{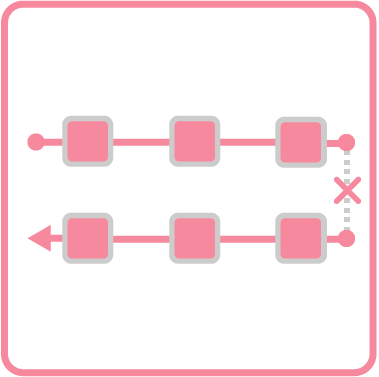}
\end{wrapfigure}
\noindent \textbf{Segmentation} enables users to split event sequences into sections, which is typically used to narrow the scope of exploration by focusing on sequence segments that are shorter than the entire sequence. Meaningful sequence segments can also indicate event occurring patterns. For example, in MAQUI~\cite{law2018maqui} and DecisionFlow~\cite{gotz2014decisionflow}, users can segment a set of event sequence by user-specified milestone events to reveal event patterns and correlations.

\begin{wrapfigure}{l}{0.042\textwidth}
\centering
\includegraphics[width=0.059\textwidth]{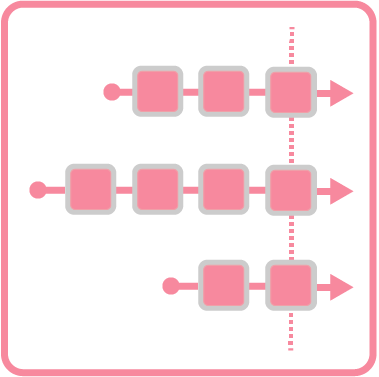}
\end{wrapfigure}
\noindent
\textbf{Alignment} refers to arranging multiple sequences to make them aligned on a selected event or time point. This interaction aims to explore and compare patterns before and after the alignment point within a single sequence or across multiple sequences. For instance, Lifelines2 \cite{wang2008aligning} supports interactive alignment of event sequences based on a selected event, so that users can easily spot precursor, co-occurring, and aftereffect events. Chen et al.~\cite{chen2018sequence} allow both sequence alignment and the adjustment of temporal scale to illustrate the temporal distribution of events with respect to a selected event. 

\begin{wrapfigure}{l}{0.042\textwidth}
\centering
\includegraphics[width=0.059\textwidth]{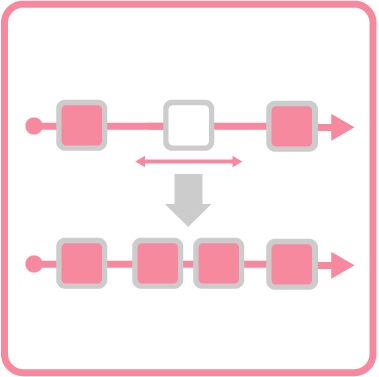}
\end{wrapfigure}
\noindent\textbf{Scaling} provides analysts an access to zoom in/out the visualizations or inspect data under various granularity. \emph{Zoom in/out} are commonly used in many visualizations, which allows a visualization-level scaling through enlarging or contracting the visual representations to enhance local details or get an overall impression. Additionally, some visual analytics techniques~\cite{nguyen2018understanding,guo2019visual,carepre,chen2018sequence} also allows a data-level scaling through \emph{abstract/elaborate} to accommodate the complexity of event sequence. For example, Guo et al.~\cite{guo2019visual,guo2018visual} allows a stage-level abstraction and elaboration by aggregating and expanding events within the same progression stage.

\begin{wrapfigure}{l}{0.042\textwidth}
\centering
\includegraphics[width=0.059\textwidth]{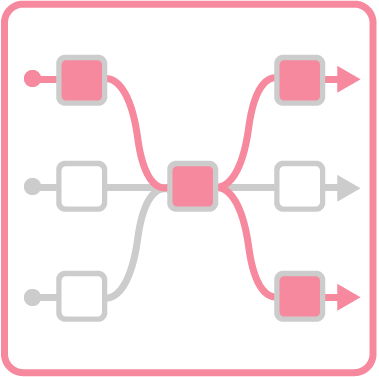}
\end{wrapfigure}
\noindent\textbf{Emphasis} aims to facilitate the discovery of interesting patterns \cite{shi2019visual}. This can be achieved through various forms of interactions such as \emph{highlighting}, \emph{sorting}, and \emph{layout adjustment}. \emph{Highlighting} draws users' attention through tweaking basic visual representations (e.g., color, size), which are commonly used in emphasizing sequence groupings, progression pathways, and critical events. \emph{Sorting} emphasizes the ranking of sequences or patterns under specific metrics. For example, Lifeflow~\cite{wongsuphasawat2011lifeflow} allows users to sort progression pathways by the number of records or average time span. \emph{Layout adjustment} enables users to arrange the positions of visual elements in a meaningful way. For example, Guo et al.~\cite{guo2017eventthread} proposed a layout algorithm that arranges sequence clusters to imply their similarities, which allows users to adjust the similarity threshold to generate different groupings.

\begin{wrapfigure}{l}{0.042\textwidth}
\centering
\includegraphics[width=0.059\textwidth]{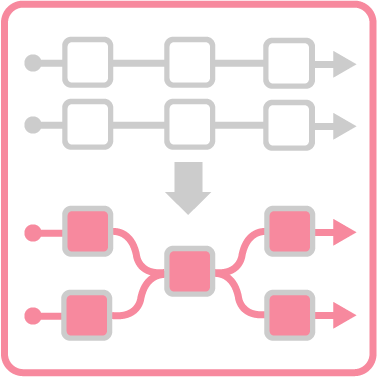}
\end{wrapfigure}
\noindent \textbf{Aggregation} enables users to interactively merge event sequences, supporting a more scalable exploration of large-scale complex event sequences. For instance, DecisionFlow \cite{gotz2014decisionflow} aggregates sequence with similar occurrence of milestone events so as to enhance visual scalability of large-scale events. CareFlow~\cite{perer2013data} merges sequences by common event occurrences to reveal frequently observed progression patterns.

\vspace{-5mm}
\subsection{Visual Analysis Tasks}
From our review of both data analysis and visual analytics techniques, we summarize the motivating analytical tasks that have gained attention from researchers over the past decade. For simplicity, we further classify these tasks into five high-level categories introduced as follows according to their fundamental objectives. 

\begin{wrapfigure}{l}{0.042\textwidth}
\centering
\includegraphics[width=0.059\textwidth]{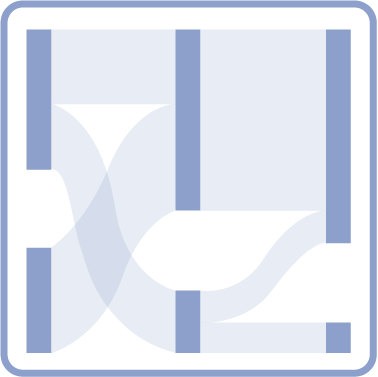}
\end{wrapfigure}
\noindent \textbf{Summarization}: Summarizing event sequences aims to uncover major progression patterns and featured groupings of the sequence entities. The fundamental motivation is to help analysts quickly get an overview of the sequence dataset. A variety of analytical tasks serve the purpose of generating summaries, including sequential pattern mining~\cite{perer2014frequence,liu2017coreflow} that discover frequently occurred subsequences from the sequence dataset, progression analysis~\cite{gotz2014decisionflow,guo2018visual} that reveals time-evolving patterns of latent progression stages, and sequence clustering~\cite{vasabi,gotz2020} that segments sequence dataset into groups.

\begin{wrapfigure}{l}{0.042\textwidth}
\centering
\includegraphics[width=0.059\textwidth]{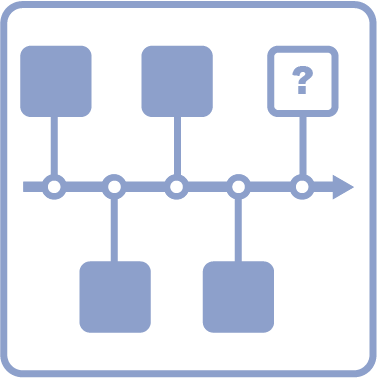}
\end{wrapfigure}
\noindent \textbf{Prediction \& recommendation}: Prediction \& recommendation tasks generally involves analyzing observed event sequences to foresee the upcoming events or sequences, or examining how certain interventions may effect the future trends. The fundamental objective is to make predictive analysis. Typical motivating tasks include making predictions on future events and outcomes~\cite{du2020interactive, carepre}, and making recommendations on user actions to help achieve certain goals~\cite{du2016eventaction}. In addition, due to the importance of interpretability in the applications of sequence predictions,
thus, we also include a group of work that visualize the underlying mechanisms of the prediction model~\cite{kwon2018retainvis} to aid result interpretation.

\begin{wrapfigure}{l}{0.042\textwidth}
\centering
\includegraphics[width=0.059\textwidth]{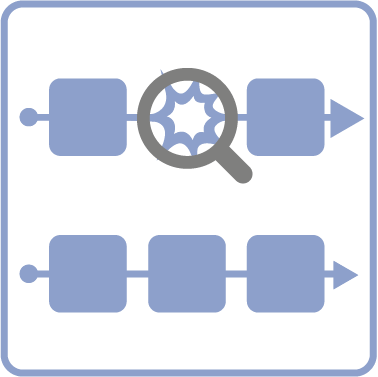}
\end{wrapfigure}
\noindent \textbf{Anomaly detection}: Visual anomaly detection for event sequences aims at identifying rare cases that deviates from the majority of the sequence progressions. Anomalies in event sequences can take multiple forms depending on the data scales (i.e., anomalous event, subsequence, sequence). For example, EventThread3~\cite{guo2019visual} detect anomalous events that derive from normal expected progressions, MOOCad \cite{mu2019moocad} identifies anomalous studying patterns of online students, and FluxFlow \cite{zhao2014fluxflow} captures anomalous spreading process of tweets. 

\begin{wrapfigure}{l}{0.042\textwidth}
\centering
\includegraphics[width=0.059\textwidth]{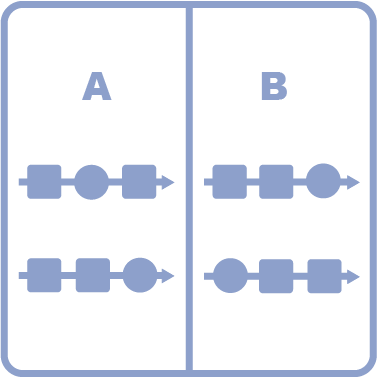}
\end{wrapfigure} \noindent
\textbf{Comparison}: Comparison is a common task when investigating similarities and differences between event sequences. Existing visual comparison techniques can be broadly categorized by the scale of comparison targets. For instance, Similan \cite{wongsuphasawat2009finding} compares individual events of two sequences, while CoCo \cite{malik2015cohort} and MatrixFlow~\cite{perer2012matrixflow} compare two collections of event sequences. 

\begin{wrapfigure}{l}{0.042\textwidth}
\centering
\includegraphics[width=0.059\textwidth]{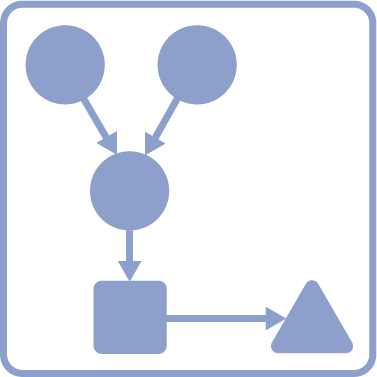}
\end{wrapfigure} \noindent
\textbf{Causality analysis}: Causality analysis aims to uncover the causal relationships between event types, promoting a better understanding of which event is very likely to occur after another, or what bring about a certain change to the outcome event. Despite that causality analysis for event sequences have gained much attention in the data mining community, the work in the field of visual analytics under this topic is still very limited, indicating a promising future research direction.  

\vspace{-3mm}
\section{Visual Analysis Techniques}
\label{sec:vis}
In this section, we summarize visual analysis techniques developed for analyzing event sequence data according to the analysis tasks introduced in Section~\ref{sec:design}.
\vspace{-3mm}
\subsection{Visual Summarization}
\begin{figure*}[t]
	\centering    
	\includegraphics[width=0.9\linewidth]{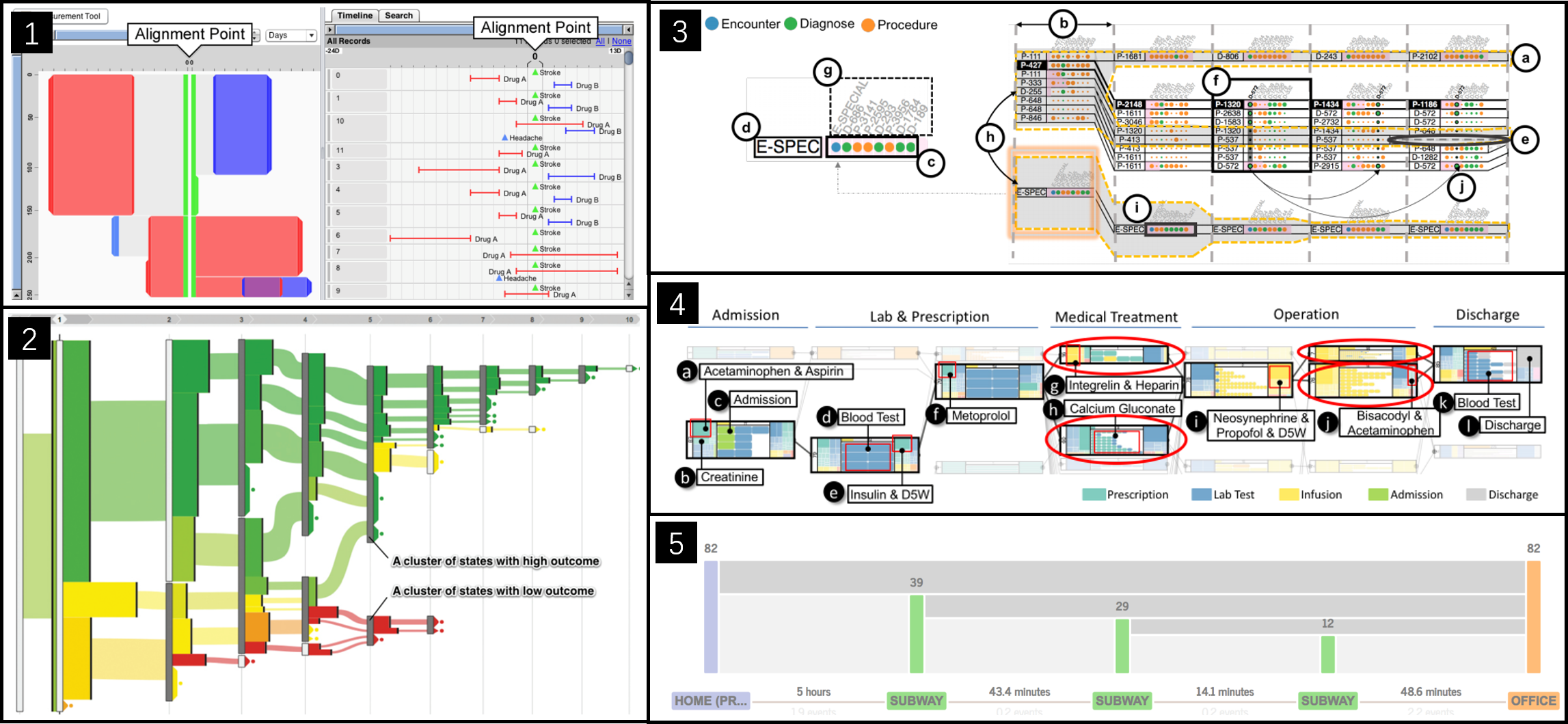}
	\vspace{-3mm}
	\caption{Selected examples of visual summarization techniques. (1) EventFlow \cite{monroe2014interactive} visualizes event sequences in both an aggregated tree-like overview and detailed a timeline display.
	(2) Outflow \cite{wongsuphasawat2011outflow}  visualizes alternative progression paths using color-coded edges that map to patient outcome. 
	(3) EventThread \cite{guo2017eventthread} visualizes the threads derived by tensor analysis as segmented linear stripes, following a line map metaphor. 
	(4) EventThread2 \cite{guo2018visual} uses a node-link visual design to provide a higher-level summary of progression patterns of event sequences.
	(5) MAQUI \cite{law2018maqui} applies a hierarchy-based visualization to represent multiple frequent patterns and adopts a timeline to reveal the temporal information.}
	\vspace{-4mm}
	\label{figure:sum}
\end{figure*}
Summarization of event sequences aims to use intuitive representations to reveal major progression patterns and featured groupings of the sequence entities.
In many domains such as health informatics \cite{plaisant1996lifelines,cappers2017exploring,han2015visual,perer2013data,Sarikaya2016,franklin2016treatmentexplorer,nielsen2009abyss}, social media \cite{perer2014frequence,law2018maqui}, and career design \cite{guo2017eventthread,guo2018visual},
a variety of analytical tasks serve the purpose of generating summaries, 
including 
\textbf{explicit summarization}, \textbf{inexplicit summarization}, \textbf{progressional analysis}, and \textbf{clustering}.
\newline\indent \textbf{Explicit summarization techniques} uncover informative patterns within event sequences using aggregated display overview. All of the sequences are visualized and aggregated into one interface. Existing techniques adopt various visualization approaches to display event sequences, such as timeline-based \cite{plaisant1996lifelines,wang2008aligning,cappers2017exploring,li2019visualizing,kiernan2009constructing,zhang2018idmvis}, sankey-based \cite{wongsuphasawat2011outflow}, and hierarchy-based \cite{wongsuphasawat2011lifeflow,trumper2012viewfusion,monroe2014interactive,Rosenthal2013} visualizations. Timeline-based visualizations are frequently adopted to reveal temporal information among sequences.
For instances,
LifeLines \cite{plaisant1996lifelines} and its variant \cite{wang2008aligning} leverage timeline-based visualizations to display the temporal distribution of events in varying time granularities. 
Sankey-based visualization are adopted to
to reveal the progression path of event over a period of time. In Outflow  \cite{wongsuphasawat2011outflow}, 
alternative clinical pathways within EMRs are visualized using a sankey diagram, the colors of path encode the  patient outcomes (Fig.~\ref{figure:sum}(2)). 
Furthermore, hierarchy-based visualizations, such as tree map and icicle plot, are able to reveal the hierarchical organizations within event sequence data.
EventFlow \cite{monroe2014interactive} reveals aggregated sequences in a hierarchy-based visualization, and individual sequences are detailed display in a list of timelines (Fig.~\ref{figure:sum}(1)). Similarly, LifeFlow \cite{wongsuphasawat2011lifeflow} leverage a hierarchy-based visualization to provide an overview of event sequences.
Explicit summarization techniques can reveal event sequences with minimum information loss, but interface will become visually messy when the scale of event sequences is large. 
\newline  \indent
\textbf{Inexplicit summarization techniques} leverage data mining metrics
to uncover informative patterns (e.g., frequent patterns) among event sequence data. 
Existing works that serve the purpose of inexplicit summarization mainly falls into two categories: query-based techniques and mining-based techniques.
\emph{Query-based techniques} \cite{krause2015supporting,zhang2015iterative,wongsuphasawat2009finding,vrotsou2009activitree,fails2006visual} enable analysts to create complex queries to extract event sequences of interest. For instances, in COQUITO \cite{krause2015supporting} and CAVA \cite{zhang2015iterative}, analysts can express complex queries for iterative cohort construction.  
\emph{Mining-based techniques} leverage
advanced sequential pattern mining algorithms to extract insights from complex event sequences
\cite{perer2014frequence,kwon2016peekquence,perer2015mining,liu2017coreflow,liu2016patterns,liu2016mining}. 
For instances, in Frequence \cite{perer2014frequence} and its variant \cite{perer2015mining}, large scale EMRs data are represented by a set of extracted frequent patterns. The authors used a sankey diagram to reveal the patterns with a color map to encode associated outcomes. 
Through this view, physicians and clinical researchers can easily understand the important correlations between treatment patterns and associated outcomes. 
But the frequent patterns do not always correspond to important or meaningful information within the data. Thus, CoreFlow \cite{liu2017coreflow} extracts branching patterns in event sequences using the Rank-Divide-Trim three-step procedure, and visualize the patterns as a tree diagram that illustrates an overview of the event flow.
\nolinebreak
\newline \indent
In addition, exploring event sequences by defining queries or using mining algorithms alone may becomes insufficient in some cases.
To this end, Law et al.~\cite{law2018maqui} proposed MAQUI, which interweaves \emph{quering} and \emph{mining} to extract informative patterns within a set of event sequences. The authors applied a hierarchy-based visualization and a timeline-based visualization to represent frequent patterns and temporal information, respectively  (Fig.~\ref{figure:sum} (5)). 
Similarly, Sequence Synopsis combine \emph{querying} and a \emph{mining} method named Minimum Description Length principle to extract informative patterns from event sequences with minimizes information loss. Each extracted patterns are visualized as a series of colored rectangles, where each rectangle represent an type of event.
\newline \indent
\textbf{Progression analysis} aims to uncover the evolution of one event during a period of time. 
Most of the aforementioned techniques produce highly summarized results, but  fail to show important low-level event details (e.g., single event features)
\cite{guo2017eventthread}. 
Visual progression analysis techniques, such as \cite{gotz2014decisionflow,gotz2016adaptive,perer2012matrixflow,guo2017eventthread,guo2018visual,chen2018stagemap}, have been introduced to reveal time-evolving patterns of latent progression stages. For instances,
in DecisionFlow \cite{gotz2014decisionflow}, analysts can use a milestone-based approach to retrieve progression patterns of interest, and visualized them in a hierarchy-based visualization. 
EventThread \cite{guo2017eventthread} has been introduced to  
summarize latent sequential patterns within a large-scale sequence collection. This technique employs a clustering algorithm to group the summarized patterns into various categories at different stages. In order to clearly reveal the summarized latent patterns, the authors adopted a line map metaphor to display the overall evolution of the latent patterns (Fig.~\ref{figure:sum}(3)). Based on preceding works, Guo et al. further proposed EventThread2 \cite{guo2018visual}, a visual analytics technique that identifies semantically
meaningful progressions using a unsupervised algorithm. This technique
solves the time scale limitation of EventThread and proposed a new
visual design to reveal the progression patterns. It combines a node-link-based cluster view and a timeline-based sequence view to provide a higher-level summary of progression patterns of multiple event sequences by grouping similar segments at each stage (Fig.~\ref{figure:sum} (4)).
\begin{figure*}[!h]
	\centering    
	\includegraphics[width=0.9\linewidth]{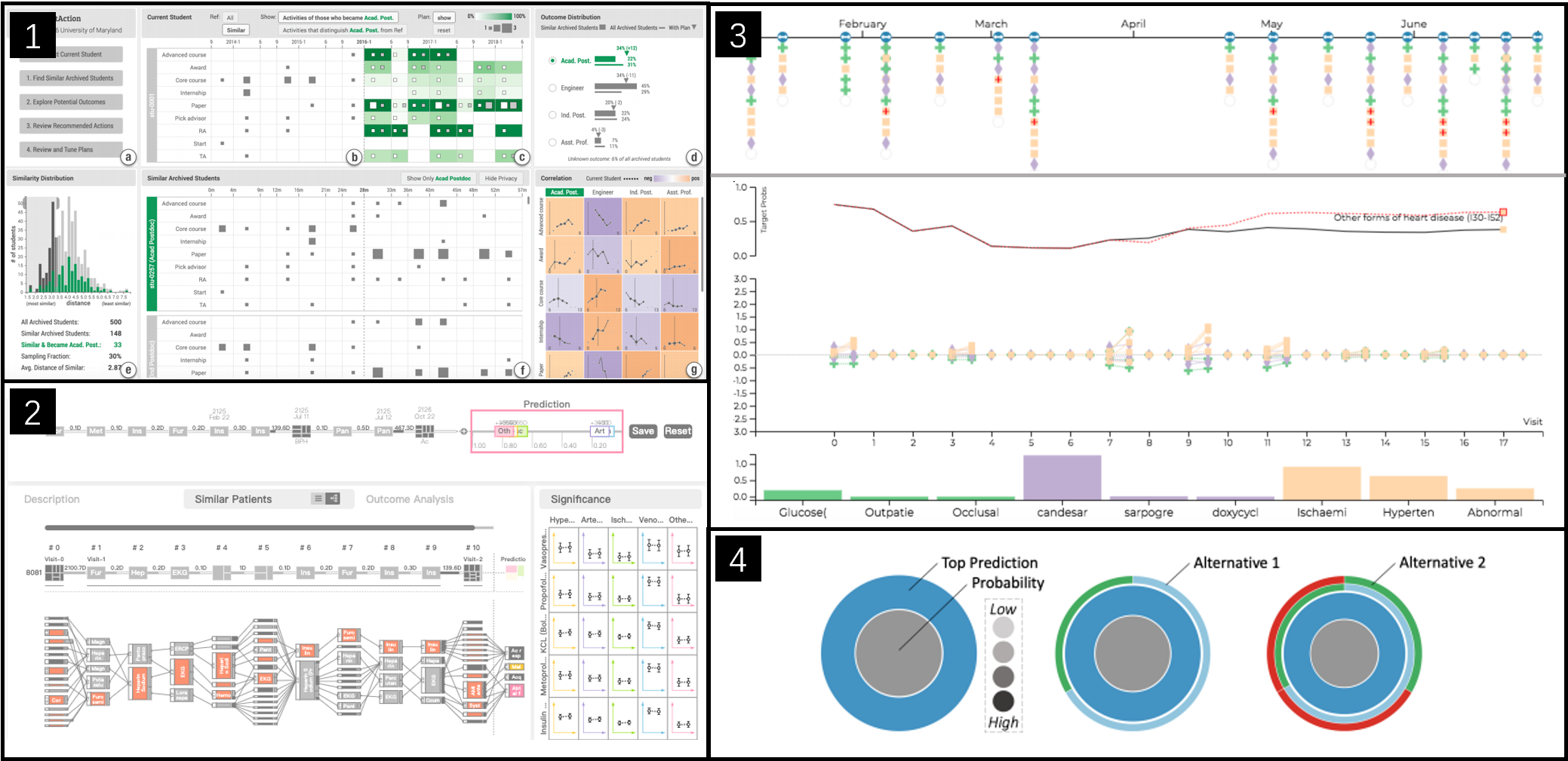}
	\vspace{-3mm}
	\caption{Selected examples of visual prediction \& recommendation techniques. (1)
	EventAction \cite{du2016eventaction} uses a calendar view to show the temporal information of event sequences. (2) CarePre \cite{carepre} reveals the medical record of a patient in a timeline-based visualization, and  similar patients' medical event sequences are aggregated into a Sankey-based visualization.
	(3) In RetainVIS \cite{kwon2018retainvis}, predicted risk trajectories are revealed in parallel line charts (middle), and the risk contributors for the patients are displayed in a bar chart (bottom). (4) \cite{guo2019visualizing} visualizes the top prediction, alternatives, and their uncertainties in a circular glyph design. } 
	\vspace{-4mm}
	\label{figure:predict}
\end{figure*}
\newline \indent
\textbf{Clustering} is the process of finding sequence-wide similarities to achieve sequence groupings.
In the clustering analysis of event sequences, a broad range of visual analytics techniques have been developed to empower analysts working with three types of event sequence data, including temporal event sequences, spatiotemporal event sequences and microarray sequences.
For temporal event sequences, clustering can be informed by sequence characteristics such as event types 
and sequence attributes. 
Cadence system \cite{gotz2020} offers a scatter-plus-focus visualization design that supports interactive hierarchical exploration of the space of event type groupings. This system adopts scented navigation cues to help users navigate complex hierarchies, as well as interactive bar charts and histograms that support additional constraints in categorical and continuous attributes of the target groups.
\cite{vasabi,rzes,wang,wei2012visual} are utilized to cluster individual entities (e.g., works, online users) based on behaviors.
VASABI\cite{vasabi} summarise user behaviours by extracting their common
tasks, and then identifies the groups of
users based on user behaviors. This technique facilitates interactive analysis of user clustering through a hierarchy-based visualization. 
DICON\cite{cao2011dicon} segments a collection of event sequences into groups based on entity attributes (e.g., age, gender). Each 
multidimensional cluster is revealed in a hierarchy-based visualization, which allows analysts to understand the event distributions in different groups.
\nolinebreak \newline \indent
The clustering analysis of spatiotemporal event sequences have been explored in many efforts, such as \cite{vrotsou,robinson,huisman2005complexities, kwan1, yu}.
Spatiotemporal visual analysis of activity diary data is visualized through VISUAL-TimePAcTS \cite{vrotsou} on a coordinate plane of time and space. Robinson et al. \cite{robinson} developed STempo, a geovisualization application to facilitate the exploration of spatiotemporal patterns within event sequence data. 
Moreover, studies have also introduced visualization tools to cluster microarray sequences \cite{hibbs, saraiya,seo2002interactively,slack2004sequencejuxtaposer}. Seo and Shneiderman \cite{seo2002interactively} created the Hierarchical Clustering Explorer that offers a dendogram and two-dimensional scattergrams, and their dynamic query controls allow users to choose which clusters to display. This model is especially suitable for bioinformatics and microarray data. Moreover, SequenceJuxtaposer \cite{slack2004sequencejuxtaposer} facilitates the comparison of biomolecular sequences using a visualization technique called “accordion drawing”. 
\newline \indent
In conclusion, visual summarization techniques save user effort by capturing
a broad view
of event sequence data. To allow for interactive exploration of visual summarization from different perspectives, the aforementioned techniques commonly employ the following interaction techniques within their interfaces: \emph{filter/query} for retrieving information of user interest, \emph{scaling} for multiple scales visualization, \emph{alignment} for aligning event sequences on selected events or time points, and \emph{sequence editing} for modifying  
event sequences during analysis.

\vspace{-2mm}
\subsection{Visual Event Prediction \& Recommendation}
Prediction \& recommendation generally involves analyzing observed event sequences to foresee the upcoming events or sequences, or examining how certain interventions may effect the future trends.
The fundamental objective is to make predictive analysis.
Typical motivating tasks include making predictions on future events and outcomes~\cite{du2020interactive, carepre,guo2020comparative,guo2019visualizing}, and making recommendations on user actions to help achieve certain goals~\cite{du2016eventaction,du2017finding}. In addition, due to the importance of interpretability in the applications of sequence predictions, a group of work that visualize the underlying mechanisms of the prediction model~\cite{kwon2018retainvis,krause2016interacting,strobelt2017lstmvis} to aid result interpretation.

\textbf{Prediction} techniques for event sequences have been proposed to predict the next event in a sequence based on historical events.
In many domains, event prediction plays an important role in decision-making. For instance, medical researchers and physicians can use this type of techniques to understand potential outcomes of patients under different treatments. 
The CarePre system \cite{carepre} leverages deep learning-based RNNs to predict the risk of a patient being diagnosed with certain diseases in the future. In this system, a patient's historical events are displayed in a timeline-based visualization (Fig.~\ref{figure:predict}(2)). Users are allowed to modify these events (e.g., by removing, moving, adding, or adjusting the event's duration) to get different outcomes
from the prediction model. Moreover, 
\cite{guo2019visualizing} is a visual analytics system designed for prediction analysis. It employs Recurrent Neural Networks (RNNs) to predict future activities, and review the most probable predictions and possible alternatives in a  circular glyph design (Fig.~\ref{figure:predict}(4)). The color of the first outer ring represents the top prediction for a group of records. Then, depending on the granularity of the analysis, alternative predictions are represented as rings and added to glyph from the inside out. 
\begin{figure*}[!t]
	\centering    
	\includegraphics[width=0.9\linewidth]{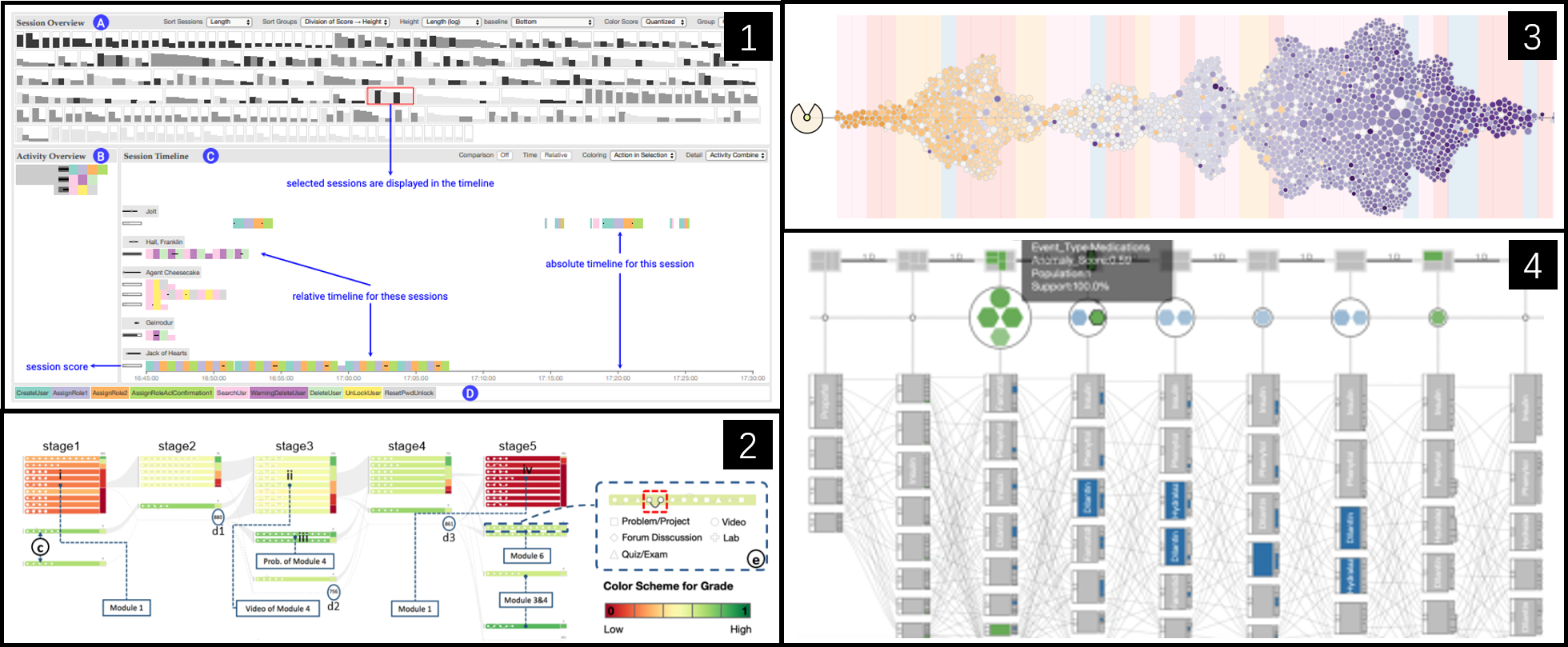}
	\vspace{-3mm}
	\caption{Selected examples of visual anomaly detection techniques. (1) \cite{nguyen2018understanding} combines a rectangle glyph design and a timeline-based visualization to reveal  anomalies within event sequences.
     (2) MOOCad \cite{mu2019moocad} employs a Sankey-based visualization to display an overview of the stage segmentation results, and it uses a matrix-based approach to indicate the content patterns of each group within the stage. (3) FluxFlow \cite{zhao2014fluxflow} visualizes anomalous retweeting sequences in a packed circles design. (4) EventThread3 \cite{guo2019visual} displays anomalous sequences in a line of rectangular nodes ordered by time of occurrence (top). The authors use a circular glyph (middle) to visualize the anomalous events within sequences.} 
	\label{figure:anomaly}
	\vspace{-4mm}
\end{figure*}
\nolinebreak \newline \indent
\textbf{Recommendation} techniques provide reliable suggestions on user actions to help achieve certain goals. Students can adopt this type of techniques to understand their future career development and find an academic plan that suits their desired goals. Du el al.~\cite{du2017finding, du2016eventaction}
introduced two career path recommendation techniques that provide suggestions and potential outcomes by summarizing the outcome of similar users. In EventAction \cite{du2016eventaction}, all of the records that similar to select one are displayed in a list of calendar views (Fig.~\ref{figure:predict}(1)). Recommendation actions are highlighted in the calendar and allow users add into their plans for next round explorations.
\nolinebreak
\newline \indent
In the past few years, deep learning algorithms have demonstrated significant improvements over traditional approaches in the tasks as prediction and classification. For event sequence data,  Recurrent Neural Networks (RNNs) are frequently adopted to foresee the upcoming events or sequences, or exam how certain interventions may effect the future trends.
However, interpretability is recognized as a primary challenge of deep learning approaches. To address this issue, recent studies have introduced visual prediction techniques to interpret the internal mechanisms of a prediction model \cite{strobelt2017lstmvis,krause2016interacting,kwon2018retainvis}. For instances, 
RetainVIS \cite{kwon2018retainvis} is a hybrid visual technique for gaining insight into how RNNs model EMR data within the context of diagnosis risk prediction tasks. This technique interprets the relationship between patient records and predicted risk scores. Specifically, patients' medical records and their predicted risk trajectory are visualized in two parallel line charts (Fig.~\ref{figure:predict}(3)), which allow users to understand the progression of predicted diagnosis risks and why such predictions are made. Also, when users hover over the x-axis, they can see the updated contribution scores of medical events, which represent the importance or contribution of an event to the predicted result. Similarly,
LSTMVis \cite{strobelt2017lstmvis} focus on the visual analysis of hidden features
in RNNs, it allows users to explore hypotheses about RNN hidden state
dynamics.
\newline \indent
In summary, visual prediction and recommendation techniques contribute to decision-making in many domains. In order to allow users to explore the data from different perspectives, the aforementioned techniques commonly employ the following interaction techniques within their interfaces: \emph{filter/query} for retrieving information of users interest, \emph{emphasis} for adjusting attributes of data to reveal interesting patterns, and \emph{sequence editing} for including a new event or a new feature into the prediction model. 
\vspace{-2mm}
\subsection{Visual Anomaly Detection}
Visual anomaly detection for event sequences aims at identifying rare cases that deviates from the majority of the sequence progressions. In many application domains, such as social media \cite{zhao2014fluxflow, cao2015episogram,cao2015targetvue}, computer systems \cite{mu2019moocad,xu2019clouddet,senin2015time}, clickstream \cite{fischer2012vistracer,nguyen2018understanding,guo2019visual}, and smart factory \cite{xu2016vidx,herr2018visual,wu2018visual}, various visual techniques have been proposed to serve the task of anomaly detection. 
As the forms of anomalies vary across different tasks,
we broadly divide existing techniques into the following three types: \textbf{anomalous events visualization}, \textbf{anomalous frequent patterns visualization}, and \textbf{anomalous sequences visualization}.
\newline \indent
\textbf{Anomalous events visualization} 
identifies anomalous events within the context of event sequences by uncovering the differences between abnormal and normal events. 
Many existing techniques incorporate multiple visualization methods in their interfaces to display anomaly events from different perspectives \cite{guo2019visual,fischer2012vistracer,nguyen2018understanding,chankhihort2017visualization,xu2016vidx,muelder2016visual}.
For instance, EventThreat3 \cite{guo2019visual} detects abnormal events within anomalous sequences based on inferred expected normal progressions (Fig.~\ref{figure:anomaly}(4)). Anomalous sequences and expected normal progressions are represented in a line of rectangular nodes ordered by time of occurrence. 
Anomalous events are revealed in circular glyphs to encode critical variables of anomalous events.
In this view, analysts can visually compare the abnormal sequence with normal sequences, and thus potentially understand why anomalies exist.
Moreover, in \cite{nguyen2018understanding},
anomalous log sequences are detected by a black-box model, and displayed in a timeline-based visualization. Each event is represented as a colored rectangle, users can verify the anomalous logs and explore the events that contribute to sequence anomaly.
Xu et al. \cite{xu2016vidx} 
extended the Marey’s graph to visualize product moving traces in a production line. The visualization of individual products and their processing times improves user understanding of a line’s performance, and also helps in better understanding anomalous events, the causes and effects in a production line.
\newline \indent
\textbf{Anomalous frequent patterns visualization} is utilized to help users perceive the anomalous frequent patterns that contribute to sequence abnormality. 
MOOCad \cite{mu2019moocad} is designed to detect anomalous learning patterns within MOOC data (a set of online learning activities sequences) (Fig.~\ref{figure:anomaly}(2)). To facilitate anomaly detection and reasoning, the large-scale learning sequences are clustered into various groups at different stages. The authors employed a sankey-based visualization to display the overview of the stage segmentation results, and a matrix-based approach to indicate the content patterns of each group within the stage. 
In this view, users can flexibly explore the anomalous learning patterns via stage comparison, group comparison within stages, and individual path inspection.
\begin{figure*}
	\centering    
	\includegraphics[width=0.9\linewidth]{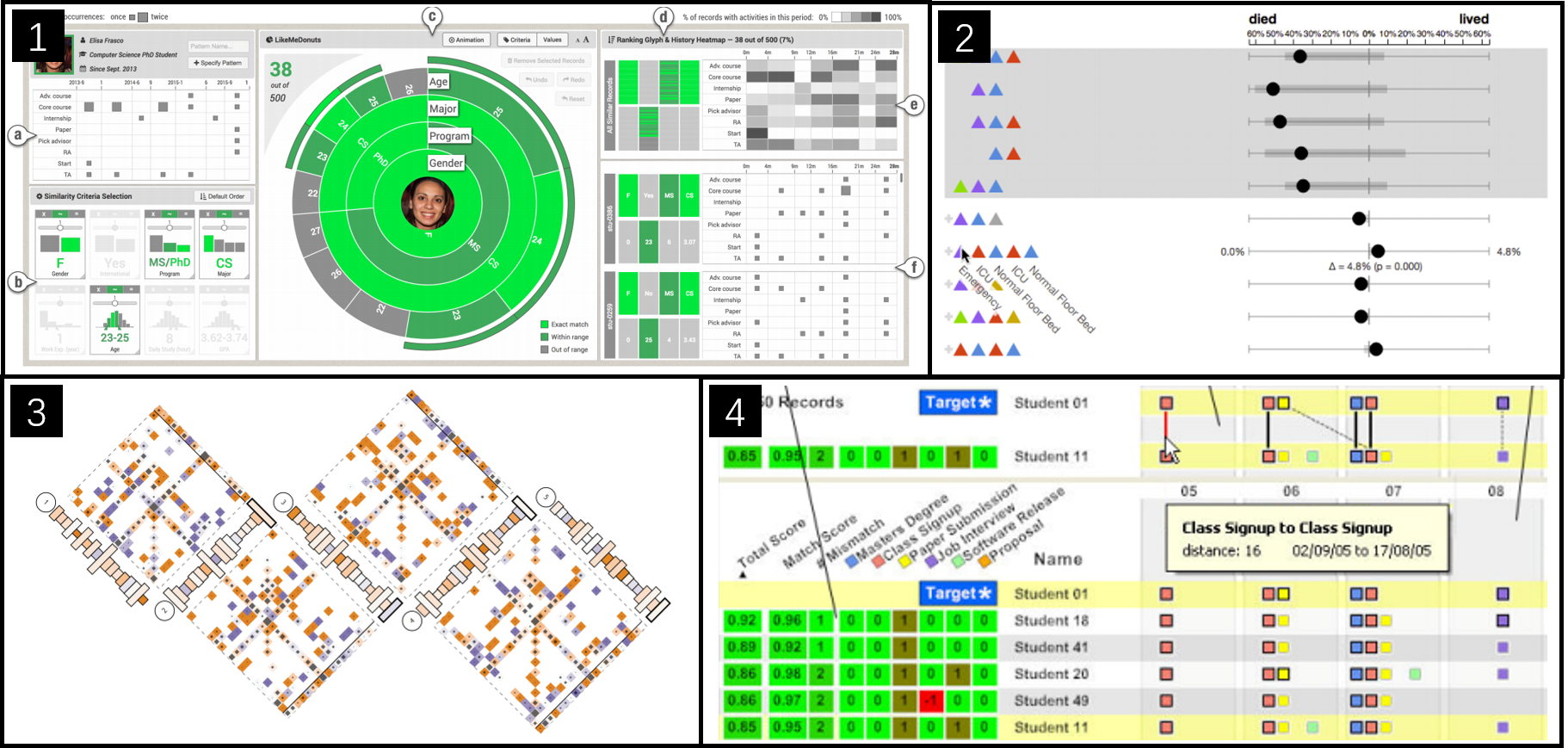}
	\vspace{-3mm}
	\caption{Selected examples of visual comparison techniques. (1) \cite{du2018visual} summarizes the criteria values of similar records as a hierarchical tree. Each of the records and common temporal patterns among the similar records are visualized in the calendar view. (2) In CoCo~\cite{malik2016high}, a combination of medical events is visualized in a timeline, where colored triangles represent medical events. (3)
	MatrixWave \cite{zhao2015matrixwave} visually compares two web clickstreams in a matrix-based visualization. (4) Similan \cite{wongsuphasawat2009finding} visualizes each event sequence in a binned timeline. The paired events are connected by lines.} 
	\vspace{-4mm}
	\label{figure:compare}
\end{figure*}
\nolinebreak \newline \indent
\textbf{Anomalous sequences visualization} helps users detect anomalous sequences within sequence collections, uncover the temporal structure of anomalous event sequences, and reveal the deviation of anomalous sequences from normal sequences. 
For instance, Zhao et al. \cite{zhao2014fluxflow} proposed a flexible timeline visualization technique to discover rumor-spreading processes between Twitter users (Fig.~\ref{figure:anomaly}(3)). The retweeting sequences are visualized by a packed circles design, where each participating user is represented as a circle. 
In order to intuitively display the abnormality of sequences, the authors designed a circular glyph for each retweeting sequence that summarizes its important aspects such as overall abnormality, contextual polarity, scale, and temporal information. Cao et al. developed TargetVue \cite{cao2015targetvue} to detect Twitter users with anomalous behaviors. This technique explores anomalous users via an unsupervised learning model and visualizes the behaviors of suspicious users in three glyphs (Fig.~\ref{figure:internet}(4)). These glyphs are designed to present the users' communication activities, features, and social interactions, respectively.
Nguyen et al.\cite{nguyen2018understanding} proposed a visual analytics approach that aims to detect unusual action sequences of users (Fig.~\ref{figure:anomaly}(1)). Every sequence is visually summarized in a compact glyph to help analysts spot anomalous sequences, the length and color saturation of glyph represent sequence length and anomaly scores respectively. Also, 
anomalous sequences are visualized in a timeline visualization, where each event is represented by a colored rectangle whose color maps to event type. Similar designs are proposed by \cite{fischer2012vistracer} and \cite{guo2019visual}. In \cite{fischer2012vistracer}, anomalous sequences are displayed in a timeline visualization with color encoding by event type. In \cite{guo2019visual}, Guo et al. employed an MDS projection to visually summarize the abnormality of a dataset and subsequently reveal sequences of interest in a timeline-based visualization. 
\newline \indent
In summary, visual anomaly detection has been introduced to solve various real-world problems across different application domains. To allow users to interactively explore data from different perspectives, the aforementioned techniques commonly employ the following interaction techniques within their interfaces: \emph{filter/query} for retrieving information of users interest, \emph{emphasis} for adjusting attributes of data to reveal interesting patterns, \emph{scaling} for multiple scales visualization, and \emph{alignment} for align sequences on selected events or time points.

\vspace{-4mm}
\subsection{Visual Comparison}
Visual comparison is a common task when investigating the similarities and differences between event sequence data.
A variety of visual comparison techniques have been proposed to solve real-world problems in many domains such as for career path analysis \cite{du2016eventaction,du2018visual}, 
clickstream analysis \cite{nguyen2018understanding,zhao2015matrixwave}, health information analysis \cite{guo2020comparative,carepre,malik2016high}, and generic purposes \cite{wongsuphasawat2009finding,guo2019visual}. In our work, we classify the visual comparison techniques for event sequences based on compared targets, including \textbf{comparison techniques for event sequences}, \textbf{comparison techniques for sequential patterns}, and \textbf{comparison techniques for sequence collection}. 
\newline \indent
\textbf {Comparison techniques for event sequences} are utilized to compare individual events in terms of disorder,
missingness or redundancy, and difference in the occurrence of timing and attributes.
To facilitate the interpretation of compared results, researchers adopted \emph{juxtaposition design} \cite{wongsuphasawat2009finding}, \emph{superposition design} \cite{carepre}, and \emph{hybrid design} \cite{guo2019visual,guo2020comparative} to clearly visualize the similarities and differences between sequences. For instance, Similan \cite{wongsuphasawat2009finding} 
shows the similarity of events within two similar sequences via \emph{juxtaposition}. In this technique, each event sequence is visualized in a binned timeline (Fig.~\ref{figure:compare}(4)), similar sequences are placed beneath the target sequence for explicit comparison. In order to reveal the similarity information between two sequences, the pairs of events matched by the Match \& Mismatch measure are connected by lines, and events without any links connected to them are missing or they are extra events.
Moreover, in CarePre \cite{carepre}, the visual comparison techniques with \emph{superposition design} are utilized to verify the predicted risk of potential diseases. Last but not least,
in the most recently published visual comparison technique, Guo et al. \cite{guo2020comparative} 
searched similar medical records and applied \emph{hybrid design} to convey the differences between target record and its similar records. This technique uses \emph{explicit encoding} to reveal the overall dissimilarity of similar records over time, and it uses \emph{superposition} to represent differences between target record and its top three similar records in detail.
\newline \indent
\textbf{Comparison techniques for sequential patterns} are used to investigate the similarity of sequential patterns within two event sequences for diverse applications, such as for log files \cite{nguyen2018understanding,qi2019stbins} and career paths \cite{du2016eventaction,du2018visual}. 
The career recommendation technique, EventAction \cite{du2016eventaction}, uses a calendar view to show event sequences and \emph{juxtaposed} them in a ranked list for visual comparison.
In \cite{nguyen2018understanding}, Nguyen et al. adopted visual comparison with \emph{superposition} to indicate the anomaly path within abnormal sequences. 
Du et al. \cite{du2018visual} supports \emph{explicit encoding} and \emph{juxtaposition} of differences for semantically meaningful comparisons (Fig.~\ref{figure:compare}(1)). Specifically, while comparing the target record with the entire dataset, the authors summarized the criteria values of the similar records in a hierarchical tree, where the similarities and differences are explicitly encoded. For a 
detailed inspection, all records and common temporal patterns are visualized in the calendar views, so that users can juxtapose any two sequences of interest to explore the differences between them. 
\newline \indent
\textbf{Comparison techniques for sequence collection} aim to find differences between two sets of event sequences in terms of structure, attribute, temporal information \cite{malik2015cohort}.
For instance, CoCo \cite{malik2016high} leveraged automated statistical analysis to compare the attributes of two distinctly defined cohorts, adopting \emph{explicit encoding} to convey an overview of the differences between the two cohorts (Fig.~\ref{figure:compare}(2)). 
Moreover, 
MatrixWave \cite{zhao2015matrixwave} is a matrix-based visualization for comparison analysis of two web clickstream datasets in terms of traffic patterns (Fig.~\ref{figure:compare}(3)). The authors applied \emph{superposition} to represent two related event sequence datasets within one visualization and used \emph{explicit encoding} to reveal the differences between traffic paths at each node. This technique focuses on differences in the occurrence of immediate and pairwise steps among two clickstream datasets.
\newline \indent
In conclusion, visual comparison techniques can save analysts' efforts to explore the differences between two event sequences or two groups of event sequences. To facilitate interactive analysis, the aforementioned techniques adopt \emph{filter/query} for retrieving information of users interest, \emph{scaling} for multiple scales visualization, \emph{alignment} for align sequences on selected events or time points, and \emph{emphasis} for adjusting attributes of data to reveal interesting patterns.

\vspace{-2mm}
\subsection{Visual Causality Analysis}
\label{sec:visual_causal}
Visual causality techniques have been proposed to help users uncover causal relationships among data.
Traditional visualizations, such as a directed acyclic graph (DAG) or the Hasse diagram \cite{koldehofe1999distributed}, can be employed to illustrate causality to a certain extent. However, they become inefficient as an increased number of variables introduces more edge crossing. Elmqvist et al. successively proposed two visual methods, Growing Squares \cite{elmqvist2003growing} and Growing Polygons \cite{elmqvist2004animated}, which enhance node representations within a DAG with color-coded squares and polygons that help provide an overview of influences on each event in large systems. They also leveraged animations to dynamically present the temporal ordering of causality. Despite that both methods are effective in uncovering the causal relationships of events, they fail to integrate causal semantics into the graph, which is important for a deeper understanding of causal structures.
\newline \indent
To incorporate additional causal semantics, Kabada et al. \cite{kadaba2007visualizing} introduced a set of animations following Michotte’s rules of causal perception \cite{michotte1963causalite} to intuitively illustrate causal strength, amplification, dampening, and multiplicity. Recent studies have invested more effort in integrating automatic causal analysis algorithms and causality visualizations into a visual analytics system to facilitate interactive causal analysis and reasoning. In \cite{chen2011data}, Chen et al. proposed a workflow for a visual causal analysis system that aims to support decision-making by providing hypothesis generation and evaluation. This leads to a number of visual analytics systems that are designed to support interactive analysis of data correlation and causation. For example, Zhang et al. \cite{zhang2014visual} introduced a visualization tool that utilized force-directed graphs to display the correlations between numerical and categorical variables in multivariate data.
Within their interface, the authors designed a slider bar that allows users to filter the edges corresponding to weak relations. 
ReactionFlow \cite{dang2015reactionflow} was developed to facilitate a better understanding of causal relationships between proteins and biochemical reactions in biological pathways. It organizes the causal pathways into a Sankey-based structure to emphasize the downstream and upstream nature of the causal relationships. It uses animation to highlight the flow of activity through a pathway. Wang et al. \cite{wang2015visual} presented a visual interface to reveal causal relationships in a force-directed graph with a color scheme design that allows analysts to edit and verify causal links according to their domain expertise. They extended this work in \cite{wang2017visual} with a path diagram visualization to better expose
% causal sequences of the variables. 
causal relationships between variables.
\newline \indent
As prior efforts mainly focus on the causal analysis of multivariate data, few techniques exist to analyze causal relationships among events in event sequence datasets. When dealing with event sequences, three major challenges need to be specifically tackled. First, the temporal nature of event progressions adds additional causal semantics, such as causal delays and causal durations, into the causation of events. Thus, this aspect raises the bar for extracting causal relationships within event sequences. Second, the high dimensionality of events and the latent structure of hierarchies in event types add complexity to the causal graph. This requires a dedicated graph layout mechanism to handle the causal complexity. Lastly, the complexity of temporal event sequences leads to difficulty in investigating event sequence collections.

\begin{figure*}[!ht]
	\centering    
	\includegraphics[width=0.9\linewidth]{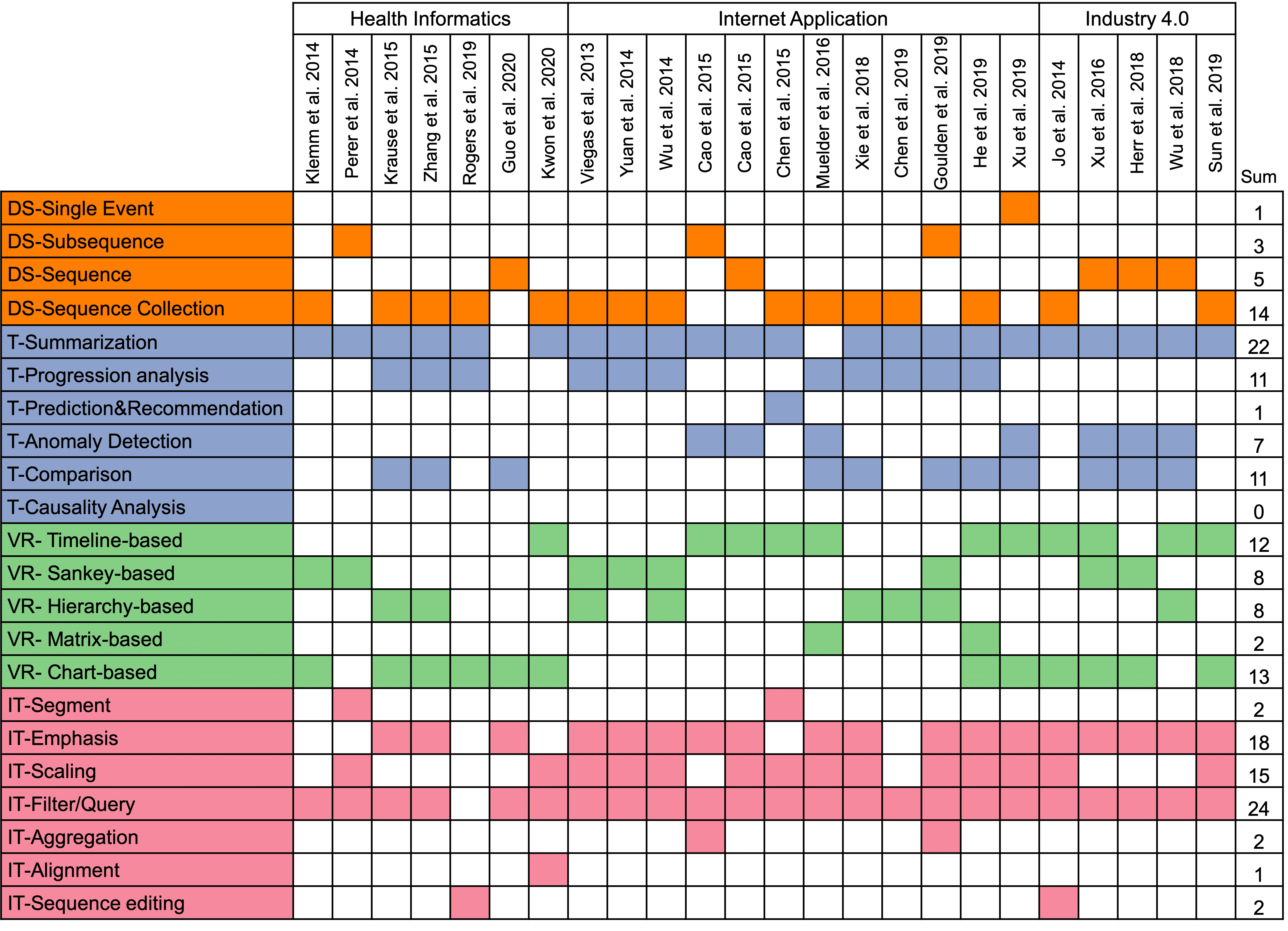}
	\vspace{-4mm}
	\caption{The selected papers regarding visualization and visual analytics of event sequences in different application domains. Each paper is labeled by relevant analysis tasks and design components in the design space. The rows are grouped and colored by dimensions of our proposed design space: DSs - Data Scales; Ts - Analysis Tasks; VRs - Visualization Representations; ITs - Interaction Techniques.} 
	\vspace{-3mm}
	\label{figure:app1}
\end{figure*}

\vspace{-3mm}
\section{Applications}
\label{sec:app}
 Given the broad applications of event sequence data, in this section, we review visual analysis techniques for event sequences applied in the fields of \textbf{Health informatics}, \textbf{Internet applications}, and \textbf{Industry 4.0}.
 
\vspace{-2mm}
\subsection{Health Informatics}
\begin{figure*}
	\centering    
	\includegraphics[width=0.9\linewidth]{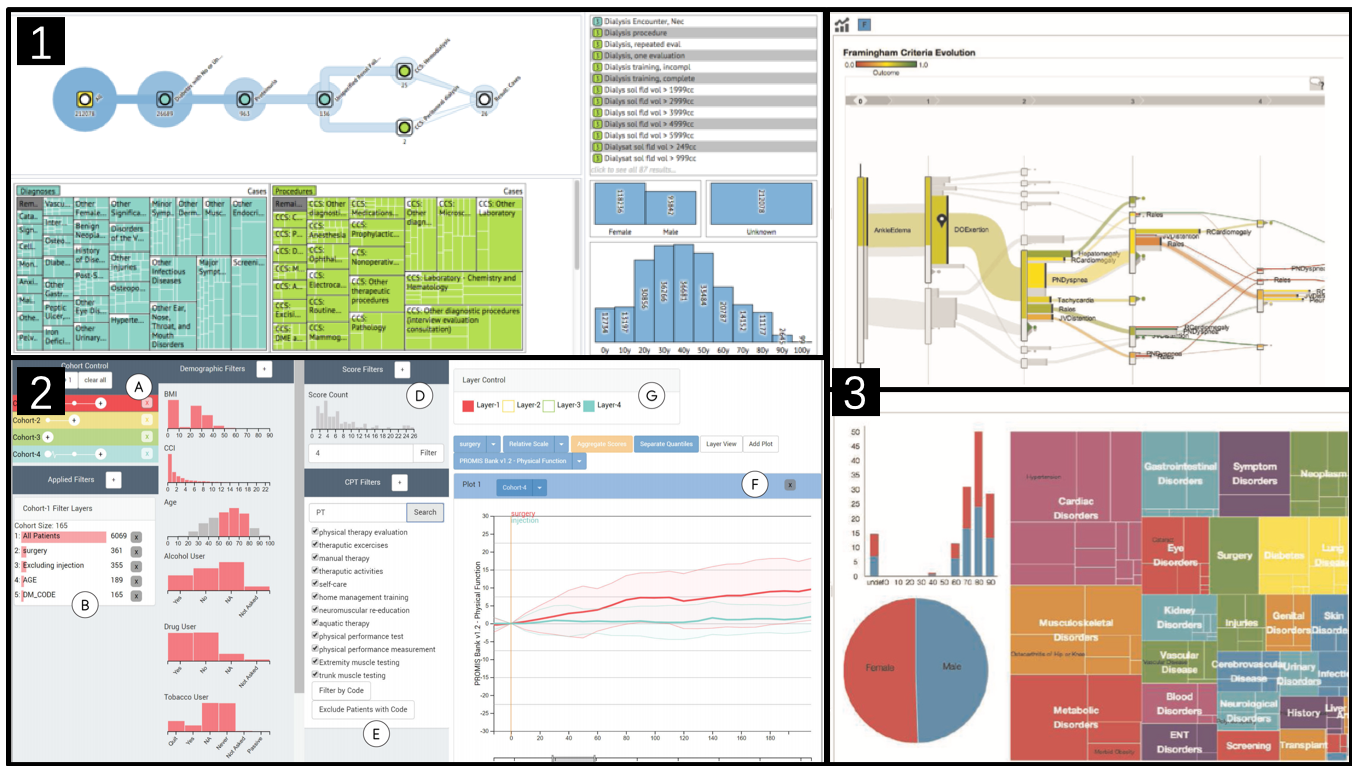}
	\vspace{-3mm}
	\caption{Selected examples of visual techniques for health informatics. (1) COQUITO \cite{krause2015supporting} uses a hierarchical tree map and bar charts to
    provide an overview of statistical information of cohorts defined by users.
    (2) Composer~\cite{rogers2019composer} plots the outcome score trajectories of different procedures in a line chart. 
    (3)
    CAVA \cite{zhang2015iterative}
    uses a stacked bar chart, a pie chart, and a hierarchical tree map to represent the
    age, gender, and diagnosis distributions of a cohort, respectively. Both the calculated risk scores and event progressions within the cohort are visualized by color-coded edges. }
    \vspace{-3mm}
	\label{figure:health}
\end{figure*}
In health informatics, electronic health records (EHRs) and electronic medical records (EMRs) can be represented as individual event sequences. Each sequence records the medical events of a patient over the course of a clinical process, and each event represents a medical event such as a diagnosis, lab test, medication, or treatment. 
With ample medical event sequence data and domain knowledge, physicians and medical researchers can extract new knowledge, quantify the effects of changes in care delivery, and potentially guide the formation of best practice guidelines.
To extract meaningful information from medical event sequences, a variety of visual techniques have been proposed to serve analysis tasks, including
\textbf{cohort analysis}~ \cite{zhang2015iterative,krause2015supporting,guo2017eventthread,guo2018visual,malik2015cohort,kwon2020dpvis,cao2011dicon}, \textbf{outcome analysis}
\cite{wongsuphasawat2011outflow,gotz2014decisionflow,wongsuphasawat2011lifeflow,wang2009temporal,rogers2019composer,perer2013data,gotz2020},
and \textbf{progonsis analysis} \cite{carepre,kwon2018retainvis}.
\newline \indent
\textbf{Cohort analysis} is a common approach used to uncover
correlations between a specific disease risk and the underlying attributes of patients within the cohorts \cite{zhang2015iterative}. 
Medical researches can 
construct a cohort of patients based on a medical event (e.g., diagnosis, treatment), the attributes of patients (e.g., gender, age), and the patterns of individual sequences (e.g., symptoms progression, treatment progression). 
Suppose a medical researcher wishes to understand the  
exposure factors for lung cancer. He can
gather the answer by analyzing common attributes within a cohort of lung cancer patients or by measuring the differences between cohorts with or without lung cancer. 
Following this idea, existing visual techniques for cohort analysis emphasize one of two strategies: cohort summarization
\cite{zhang2015iterative,guo2017eventthread,guo2018visual,polack2018chronodes}, or cohort comparison \cite{malik2015cohort,kwon2020dpvis,polack2015timestitch,krause2015supporting,cao2011dicon}.
\nolinebreak
\newline \indent
Cohort summarization techniques, such as CAVA \cite{zhang2015iterative} and Chronodes \cite{polack2018chronodes}, visually summarizing informative patterns within a cohort and uncover the common exposure factors for a disease.
CAVA \cite{zhang2015iterative}
combines chart-based and hierarchy-based visualization to represent the attribute distributions of a cohort (Fig.~\ref{figure:health}(3)). Then, to further investigate exposure events in the cohort, each patient was assigned a hospitalization risk score based on their medical history. Both the calculated risk scores and event progressions are visualized by color-coded edges, analysts can intuitively understand how different
event progression pathways lead to different hospitalization risk scores and which medical events have higher risks. Moreover, in EventThread2 \cite{guo2018visual}, the clustered medical event sequences and common sequential patterns (e.g., typical care plans) of a cohort are visualized in a sankey-based visualization and a node-link visual design respectively (Fig.~\ref{figure:sum}(4)). 
User can inspect common sequential patterns of a cohort with the goal to explore those medical events that affect further progression.
\nolinebreak \newline \indent
Cohort comparison measures differences between two cohorts of patients to
determine exposure factors of a condition such as disease or death. 
COQUITO \cite{krause2015supporting} helps users interactively construct two cohorts and explore exposure events for a disease. It uses a hierarchical tree map and multiple bar charts to
provide an overview of statistical information about the cohorts (Fig.~\ref{figure:health}(1)). 
Then it leverages PARAMO \cite{ng2014paramo}
to compare two cohorts and determine if the constructed cohorts carry exposure events for a disease.
CoCo \cite{malik2015cohort} is a visual comparison technique (Fig.~\ref{figure:compare}(2)) that measures the differences between two cohorts under various differentiating metrics. Users can select metrics of interest, such as the most differentiating event subsequences between two cohorts, to explore the medical events or patterns that may influence the incidence of a condition. In CoCo, 
each row displays the difference between two cohorts, where medical patterns of cohorts are visualized by a timeline-based design.
A circle marker is placed horizontally between two cohorts to display the difference between the values and in the direction of whichever cohort’s value (e.g., death rate, survive rate) is higher.
\newline \indent
\textbf{Outcome analysis} studies the end results of different medical progressions (e.g., symptom progressions, treatment progressions) with the goal of 
facilitating informed decision-making about diagnosis and treatment options. Existing works, such as Outflow \cite{wongsuphasawat2011outflow} and Frequence \cite{perer2014frequence}, reveal the medical progression paths in a sankey-based visualization to uncover the outcomes of different procedures. More specifically, 
Outflow \cite{wongsuphasawat2011outflow} aggregates medical event sequences from a cohort of patients and visualizes alternative progression paths using color-coded edges that map to patient outcomes (Fig.~\ref{figure:sum}(2)). Similarly, in a series of efforts proposed by Perer et al. \cite{perer2013data,perer2014frequence,perer2015mining}, 
the authors 
extracted frequent progression pathways of a cohort and 
used a sankey-based visualization to display them, 
while providing context on which care plans were successful and which were not.
These techniques provide an overview of the progression pathways within a cohort, and thus help users understand which factors, medical pathways, or other structures are most associated to the outcome of interest. 
Nevertheless, as users are not allowed to interactively build the cohorts in some outcome analysis techniques, the analytic capability of these techniques could be hugely impacted when analyzing a sequence collection of different patients. 
To overcome this issue, 
DecisionFlow \cite{gotz2014decisionflow} 
leverages a milestone approach to support users in defining a cohort by highlighting patients with a specific outcome (e.g., a disease).
In this technique, the author
used a hierarchy-based visualization to interpret how many patients within the cohort have the
specific outcome. Users can interactively compare the proportion of patients across different medical procedures and explore the association between medical events and outcomes.
Moreover, Composer~\cite{rogers2019composer} enables users to interactively explore the outcomes under different cohorts and treatment plans. This technique
employs PROMIS (Patient-Reported Outcomes Measurement Information System) to automatically evaluate the outcome scores of a patient under user-defined treatments, and plots the outcome score trajectories in a line chart (Fig.~\ref{figure:health}(2)). Medical researches can plot outcome trajectories of different treatments in one chart to determine the optimal treatment for a cohort of patients.
\newline \indent
\textbf{Prognosis analysis} predicts the
risks of a patient being diagnosed with certain diseases in the future based on the patient's medical history. A series of deep learning–based visual prognosis techniques, such as \cite{kwon2018retainvis, carepre,krause2016interacting,choi2016retain}, have been introduced to serve prognosis analysis and interpret the results.
For instance, \cite{kwon2018retainvis,krause2016interacting} implement RNNs to predict the current and future states of a patient. 
RetainVis \cite{kwon2018retainvis} enables users to modify individual sequences of medical events (e,g., add or remove medical events, modify visit period) to experiment with how predicted risk changes according to event sequences changes. 
The authors visualized a patient's predicted risk trajectory and their medical event sequences in two parallel line charts (Fig.~\ref{figure:predict}(3)). 
In this view, users are able to observe correlations between medical event sequences and prediction risks, and thus understand why such risk predictions are made. 
CarePre system \cite{carepre} can predict the risk of a patient being diagnosed with a certain disease and estimates the most influential treatments for a patient based on historical medical records. The patient's historical events are visualized in a timeline-based visualization (Fig.~\ref{figure:predict}(2)), and users are allowed to modify these events (e.g.,
removing, moving, duration adjustment, adding) to get different predicted risks. Clinicians can create multiple edited sequences to analyze the predicted results under alternative treatments, and this system thus helps clinicians understand the impact of different treatment options.

\vspace{-3mm}
\subsection{Internet applications}
\begin{figure*}
	\centering    
	\includegraphics[width=0.9\linewidth]{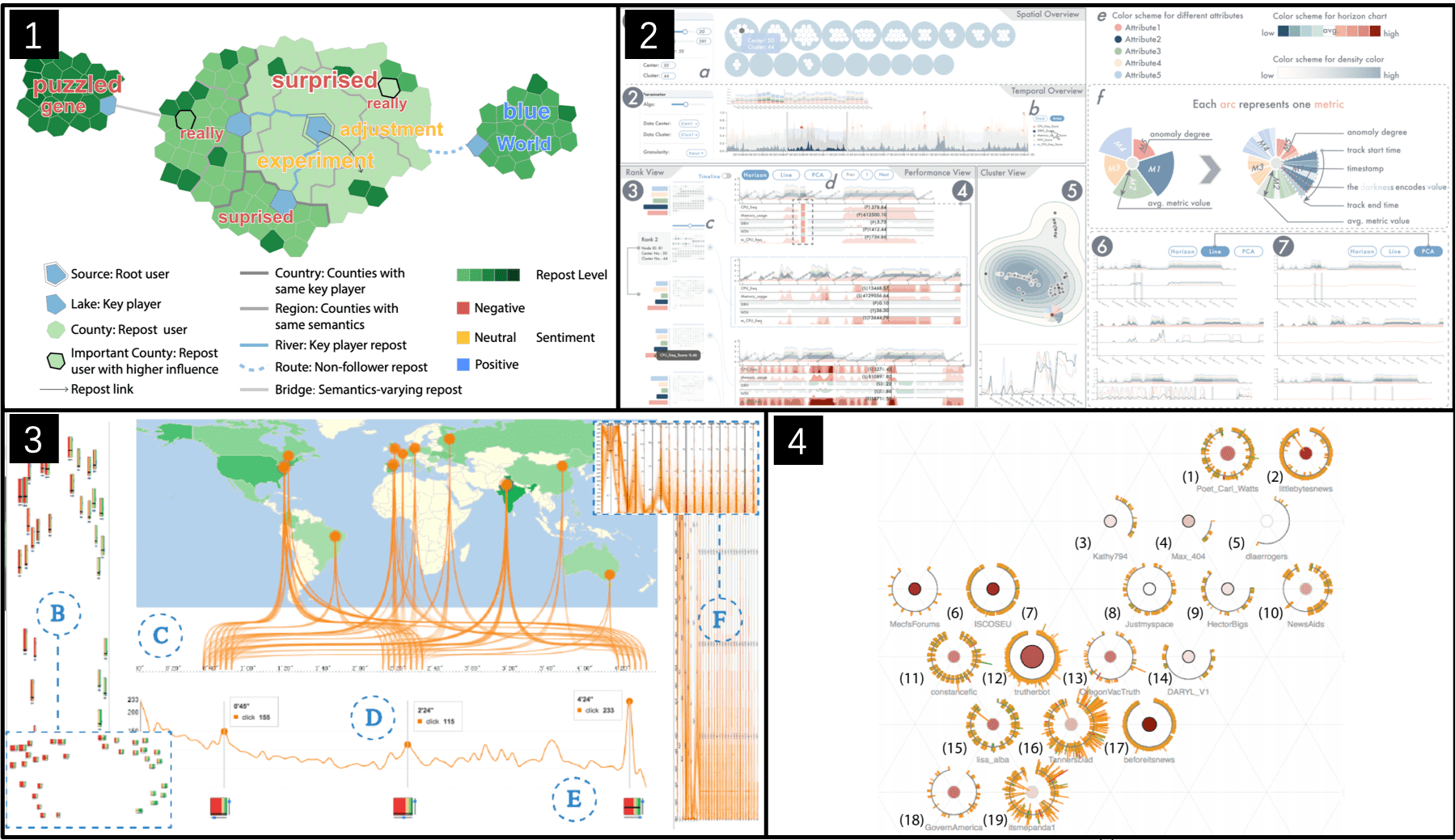}
	\vspace{-3mm}
	\caption{Selected examples of visual techniques for Internet applications. (1) R-Map \cite{chen2019r} uses a map metaphor to symbolize the reposting process in a spatial context. (2) CloudDet\cite{xu2019clouddet} combines a glyph design and a stacked line chart to monitor the performance of a computer system. (3) PeakVizor \cite{chen2015peakvizor} encodes each interaction peak by a glyph in an overview, and the spatial-temporal information of the peaks and the correlation between the peaks are visualized in two additional views. (4)
	TargetVue \cite{cao2015targetvue} visualizes the behaviors of suspicious users in three glyphs that present the user’s communication activities, features, and social interactions respectively.
	}
	\vspace{-3.5mm}
	\label{figure:internet}
\end{figure*}
In various internet applications, the activities of users and devices can be recorded as individual event sequences. For instance, social media data contain sequences of timestamped activities (e.g., posting or commenting) for specific users that are recorded over time. Similarly, clickstream data collected from e-commerce websites record how visitors operate and navigate through a web site, and this data can be represented as sequences of timestamped events (e.g., visiting a product page, purchasing a product) generated by visitor actions. Additionally, the system logs collected from a computer system can also be represented as temporal event sequences of device conditions (e.g., usage, temperature, workload).
In this section, we provide a review of the visual techniques that have been developed for event sequence data retrieved from \textbf{social media platforms}, \textbf{e-commerce websites}, and \textbf{computer systems}.

\vspace{-2mm}
\subsubsection{Social media}
On social media platforms such as Twitter and Facebook, user activities can be recorded as event sequences. Each sequence records the temporal activities of a user over time, where each event represents an online activity such as posting or commenting. 
Analysis of such event sequence data has exhibited potential for understanding various types of user behavior on social media. Existing efforts have proposed a range of visual analytics techniques to help yield insights about \textbf{collective behaviors} \cite{viegas2013google,zhao2014fluxflow,perer2014frequence,law2018maqui,yuan2014visualization,wu2014opinionflow,chen2019r} and \textbf{ego-centric behaviors} \cite{cao2015episogram,cao2015targetvue,chen2018d}.
\newline \indent
\textbf{Collective behaviors} refer to activities conducted by a temporary and unstructured group of people. On social media, collective behaviors are formulated by groups of social media users through 
the processes of spreading information and human mobility.
To study these collective behaviors and identify behavioral patterns, various visual analytics techniques have been proposed: \cite{viegas2013google,chen2019r,zhao2014fluxflow,cao2012whisper,sun2017socialwave} are designed to study the behavior of \emph{spreading information}, and  \cite{perer2014frequence,law2018maqui} can be utilized to analyze \emph{human mobility}. \newline \indent \emph{Reposting process} refers to how information spreads across space and time on social media platforms. Google $+$ \cite{viegas2013google} interweaves node-link diagrams and circular map metaphors to visualize message spreading paths. Analysts can easily capture the traces of diffusion between users and identify the importance of a message by its size and diffusion path. 
Chen et al. \cite{chen2019r} used a map metaphor to symbolize the reposting process in a spatial context (Fig.~\ref{figure:internet}(1)). 
The diffusion structure is visualized using various link metaphors such as rivers, routes, and bridges. 
This technique highlights the influence of key players, and it enables analysts to explore how these key players promote the evolution of topics and enlarge the influence of the source message.
Zhao et al. \cite{zhao2014fluxflow} proposed a flexible timeline visualization technique to reveal the rumor spreading process among Twitter users. 
Moreover, tracing the spatiotemporal information of diffusion pathways can uncover how information is spread on a global scale, such as \cite{cao2012whisper, sun2017socialwave}. Cao et al. \cite{cao2012whisper} visually summarized the temporal trends, the social-spatial extent, and community response to a topic using a sunflower metaphor. The original tweets are placed at the center of the circle and linked with geo-groups (users from a same country) once the original tweets are reposted by users in these groups. The retweeting activities are displayed as a sequence of color-coded retweet glyphs moving along pathways that indicate the timing and sentiments of the retweets.

Beside the reposting process, 
another collective behavior of importance is \emph{human mobility}. The spatiotemporal event sequences retrieved from social media platform, like Foursquare, have recently been used to uncover user mobility patterns and predict mobility decisions. For example, by studying human mobility, advertisement companies can explore the mobility patterns of people such as when and where they go to work and, thus, optimize their advertising strategies.
Some visual analytics techniques that leverage pattern mining algorithms have been used to explore common mobility patterns of users such as  \cite{perer2014frequence,law2018maqui}. MAQUI \cite{law2018maqui} support interactive exploration of the data collected from Foursquare to uncover the frequent mobility patterns of users. 
\newline \indent
\textbf{Egocentric behaviors} refer to activities conducted or influenced by a user. An egocentric perspective enables a closer analysis of individual behaviors and thus provides more detailed behavioral patterns \cite{chen2018d}. For instance,
Cao et al. \cite{cao2015episogram} proposed Episogram, an egocentric representation for visualizing individuals' interaction histories (e.g., posting or reposting content). Episogram visualizes each interaction thread using a vertical line on a timeline and uses a glyph design to represent interaction events among users. Building upon preceding works, 
Cao et al. developed TargetVue \cite{cao2015targetvue} to detect and visualize users with anomalous behaviors on Twitter. TargetVue detects anomalous users via an unsupervised learning model and visualizes the behaviors of suspicious users in three glyphs that represent the user’s communication activities, features, and social interactions, respectively (Fig.~\ref{figure:internet}(4)). Moreover, \cite{chen2018d} proposed a map-based visual technique to summarize the historical diffusion traces initiated by a central user. Users who participated in reposting one central user’s post are visualized as hex nodes whose color and size encode the user's behaviors and roles. 
These users are grouped into different regions on the map and linked with the central user, forming the social network of the central user. In this view,
if one user leads to a great amount of reposting, analysts can understand how information reaches him and diffuses from him.

\vspace{-2mm}
\subsubsection{E-commerce}
Clickstream data collected from e-commerce websites record how visitors operate and navigate through web sites. A visitor online activity can be recorded as an event sequences, in which each event represent a single online activity (e.g., visiting a product page). The increasing availability of such event sequence data permits analysts to extract valuable insights in website design and commercial activities such as advertising.
Existing visual techniques have been introduced to explore \textbf{frequent visiting traces} \cite{zgraggen2015s,liu2016patterns,liu2017coreflow} and \textbf{user behavior patterns} \cite{chen2015peakvizor,mu2019moocad,he2019vuc,goulden2019ccvis,fan2017personal}.
\newline \indent
To facilitate the understanding of frequent visiting traces, Zgraggen et al.~\cite{zgraggen2015s} proposed (s$|$qu)eries to visualizes regular patterns of clickstream data. Moreover,
Liu et al.~\cite{liu2016patterns} extracted frequent browsing paths from clickstream data and visualized them in a funnel-based visualization. As frequent patterns do not always correspond to important or meaningful information within data, CoreFlow \cite{liu2017coreflow} leveraged a tree-based visualization to facilitate branching pattern exploration for browsing paths.
\newline \indent
Analyzing clickstream data can help e-commerce companies explore users behavior and optimize their business plans. 
This idea has been extended to online education platforms with the goal to explore student learning behaviors  \cite{chen2015peakvizor,mu2019moocad,he2019vuc,goulden2019ccvis}. For instance, 
PeakVizor \cite{chen2015peakvizor} analyzes students' interaction activities to understand how students respond to video material. For example, if an unexpectedly high occurrence of pausing or 
rewinding
is observed at a certain segment, then this segment is probably difficult or confusing and thus requires additional time watching and studying. The authors encoded each peak, representing high pausing or rewinding activity, using a glyph in an event sequence overview. Moreover, spatiotemporal information about the peaks and the correlations between peaks are also visualized in two additional views (Fig.~\ref{figure:internet}(3)).
CCVis \cite{goulden2019ccvis} explores the patterns in online students' clicking behavior, and thus, identifying the course resources that were clicked most and least. It visualizes the critical sequences that lead to different transition probabilities in a node-link diagram, and it use a sankey diagram to display the click behavior patterns. 
\begin{figure*}
	\centering    
	\includegraphics[width=0.9\linewidth]{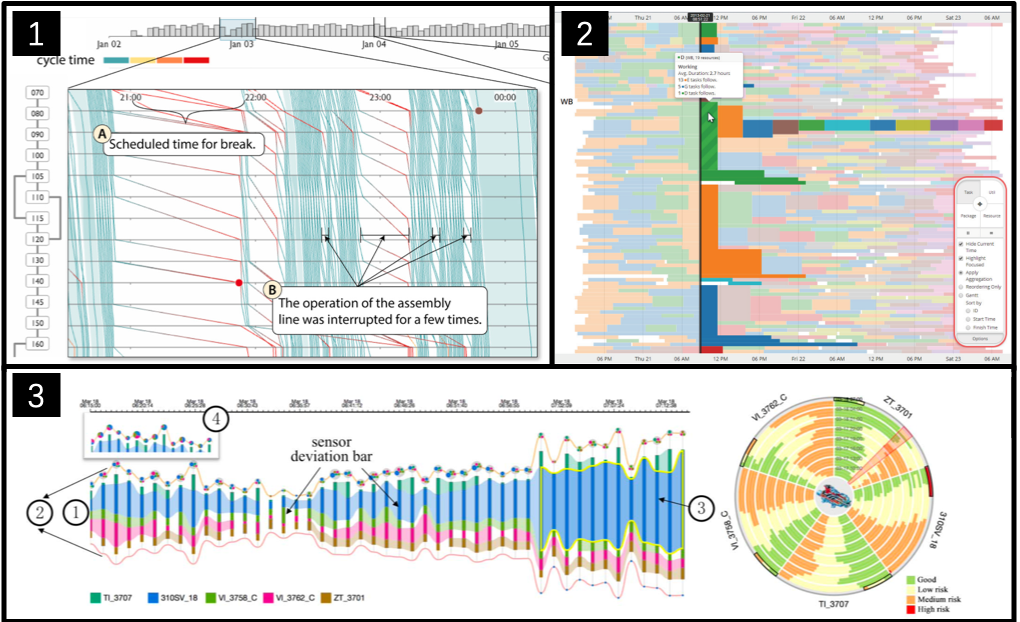}
	\vspace{-3mm}
	\caption{Selected examples of visual techniques for Industry 4.0. (1) Xu et al. \cite{xu2016vidx}
	extend a Marey’s graph to visualize product moving traces in a production line. (2) In LiveGantt \cite{jo2014livegantt}, the big picture of the current schedule is visualized in a Gantt chart. (3) Wu et al. \cite{wu2018visual} employ a stacked timeline to reveal how the real equipment conditions deviate from "normal" conditions in a period of time. The long-term trends of equipment conditions are visualized as a radial visualization to provide users with an overview of equipment conditions during a certain past time period.}
	\vspace{-4mm}
	\label{figure:industry}
\end{figure*}
\vspace{-3mm}
\subsubsection{Computer systems}
Computer systems are monitored by regularly sampling profile data that record the timestamped conditions (e.g., CPU load, memory usage) of specific devices over time as event sequences.
Monitoring and analyzing the
profile data are important for identifying devices that are over- or under-allocated, inefficient operations, and nodes that are misbehaving or failing. Muelder et al. \cite{muelder2016visual} proposed a visual technique to portray the behavior of cloud computer systems over time. The authors adopted a stacked graph timeline to summarize the aggregate behavior of cloud computer systems. For detailed inspection, 
the behavioral lines of each compute node are plotted in a table of line charts. In this view, analysts can efficiently explore the trends and anomalies within a system. Xie et al. \cite{xie2018visual} leveraged one-class support vector machines to detect anomalous executions in high performance
computing clusters. Detected anomalies are visualized in a multi-level visualization system for deeper analysis. Specifically, all of the anomalous compute trees are identified in a scatter plot. Analysts can select the anomalies of interest to inspect their structural patterns in a node-link diagram and their invoked functions in a stacked timeline. \cite{xu2019clouddet} provides interactive visualization capabilities that enables analysts to inspect profile data and identify anomalous performances in cloud computing systems. This system combines multiple visualization modes such as glyph design and stacked line charts, to comprehensively monitor the performance of cloud computing systems from different aspects (Fig.~\ref{figure:internet}(2)).

\vspace{-3mm}
\subsection{Industry 4.0}
In smart factories, the temporal status of equipment over time can be recorded as an individual event sequences, where each event represent a status (e.g., a equipment condition or a processing event). Monitoring and analyzing these event sequence data can help managers understand factory conditions, quickly respond to various sorts of events, and optimize the productivity of factories.
A variety of visual techniques have been introduced to help users exploring anomalous events \cite{herr2018visual,xu2016vidx,chankhihort2017visualization,wu2018visual,zhou2018visually} and optimizing manufacturing plans \cite{sun2019planningvis,jo2014livegantt}.
\newline \indent
In smart factories, an anomalous event (e.g., equipment failure, outlier process) could result in a serious incident or great financial loss. Traditional anomaly detection depends on manually checking every equipment, which is too expensive and inefficient. In contrast, the collected manufacturing data provides more reliable resource for factory managers to analyze anomalies. For instance, 
%In order to help managers efficiently explore the anomalies,
Herr et al. \cite{herr2018visual} analyze event reports of a production line and detected systematic issues in manufacturing processes. Reported events are shown as a time series plot that can help understand the error distribution and recurring error patterns. 
Xu et al. \cite{xu2016vidx} 
extended the Marey’s graph to visualize product moving traces in a production line (Fig.~\ref{figure:industry}(1)). 
The visualization of individual products and their processing times improves user understanding of a line’s performance, and also helps in better understanding anomalies, the causes and the effects in a production line.
The visual technique proposed by Wu et al. \cite{wu2018visual} provides an interactive interface to monitor the status of equipment in smart factories. 
The authors estimated normal conditions of equipment based on a training set, and then employed a stacked timeline to reveal how the real equipment data deviate from estimated normal conditions over a shot period of time (Fig.~\ref{figure:industry}(3)). Moreover, in order to visually summarize the long-term trends of equipment conditions, the authors adopted a radial visualization to provide an overview of equipment conditions during a certain past time period. 
\newline \indent
Analyzing manufacturing data can help managers and factory planners optimize manufacturing schedules. More specifically, in a production line, each machine is responsible only for a specific part of the production process. When the cooperation of machines is not well designed, the production line's overall efficiency will be negatively affected. The event sequence data of production lines record the past and current tasks of machines. By analyzing these data, factory planners can explore and reschedule inefficient plans, such as a manufacturing plan with significant equipment conflict. LiveGantt \cite{jo2014livegantt} is an interactive schedule visualization tool that helps managers explore highly concurrent manufacturing schedules from various perspectives. In this technique, the big picture of the current schedule is visualized in a Gantt chart (Fig.~\ref{figure:industry}(2)). Users are allowed to interactively explore the inefficiencies and reschedule manufacturing plans accordingly. 
PlanningVis \cite{sun2019planningvis} is a multi-level visualization system to support interactive exploration and comparison of production plans. This technique juxtaposes heat maps, line charts, and bar charts to visualize the differences between two plans, and thus, optimizing production plans.

\vspace{-4mm}
\section{Challenges and Opportunities}
\label{sec:discussion}
In previous sections, we summarized event sequence visualizations according to our proposed design space, extracted five analytical tasks common in visual analysis techniques for event sequences, and categorized the visual analysis techniques into three typical applications. Through this process, we found several remaining
challenges in existing research and promising future research directions that are discussed in this section.

\textbf{Data quality:} The performance of data analysis techniques largely depends on the quality of data~\cite{kandel2011research}. On top of this, the complexity of event sequence data adds difficulty in data recording and leads to more problems for data quality. Typical data quality issues include data incompleteness (e.g., missing events or timestamps), data errors (e.g., errors or inconsistency in event naming), and duplication of data records, each which can mislead statistical analysis results. The issue of data quality implies a need for additional effort to improve data processing to prevent misleading results and inferences gathered from the source data. 

\textbf{Uncertainty:} Uncertainty in information is introduced when analyzing event sequence data with quality issues or during user-specified data adjustments such as data transformation and wrangling. This uncertainty can inhibit analysts from making optimal decisions if information about uncertainty is not properly communicated in the visual analytics process~\cite{sacha2015role}. Although some previous studies~\cite{guo2019visualizing, du2020interactive} have incorporated uncertainty information in visual analytics of event sequence data, they focused on only one type of uncertainty information -- the probabilistic uncertainty under an event prediction scenario. 
Therefore, more research is required to study the best ways of incorporating and visualizing other types of uncertainty information, such as bounded uncertainty, during the process of event sequence data analysis.

\textbf{Scalability:} Scalability is a well-recognized challenge in visual analytics~\cite{keim2008visual,cook2005illuminating}. This problem becomes more significant in visual analytics of event sequence data due to the large scale (i.e., large number of sequences) and high dimensionality (i.e., vast number of event types) of most real-world event sequence datasets~\cite{gotz2014decisionflow}. Some previous research touches upon this problem mainly through sequence aggregation~\cite{wongsuphasawat2012exploring} and event filtering~\cite{gotz2014decisionflow,gotz2020} to enhance the visual scalability on the sequence level and event level respectively. However, these summarization techniques hinder the inspection of detailed individual sequences and events, and the problem of how to scale across both sequence summarizations and low-level details still remains. Therefore, there 
is a demand for a scalable visual analytics pipeline that follows the Visual Information-Seeking Mantra
by Shneiderman~\cite{shneiderman1996eyes}: ``overview first, zoom and filter, then details-on-demand'' to allow users to flexibly switch between visual summaries and sequence details.

\textbf{Heterogeneity:} Event sequence data can contain a variety of heterogeneous temporal events. For example, medical health records usually include multiple event types such as diagnostic events, lab test results, vital signs, drug administrations, etc. Events of each event type are observed or recorded with different sampling rates and show different event patterns, which leads to great difficulty for aggregating and organizing data from multiple sources. Most existing techniques choose to assemble all types of events to form a unified process for modeling and display. However, this may hinders the discovery of relationships between event types and distinctive patterns from disparate event types, which is crucial for investigative tasks and sense-making processes~\cite{wang2009visualization}. To 
solve this issue, a visual analytics framework need to be developed, enabling both the integrated analysis of multiple event processes and the investigation of patterns and trends for individual processes.

\textbf{Multivariate event sequence visualization:} Existing visual analytics techniques for event sequences generally characterize events based on their types and timestamps only. Besides these two common event attributes, however, events in a sequence can also be associated with multivariate data. For instance, lab test events in medical data are associated with specific test values, and financial transaction records also contain information about bank accounts and the monetary amount of a transaction. It still remains challenging to visualize multivariate event sequences due to the large number of event attributes a single event may include, coupled with the additional heterogeneity introduced by different data formats of the variables linked to events. Cappers and Wijk~\cite{cappers2017exploring} provide a starting point of this issue by displaying the distributions of attributes for each individual event using lists of bar charts. However, this method can be limited for the discovery of association between attributes of the same event or between multiple event types. This implies a need for a new visualization design that is able to represent categorical event types and multivariate attributes 
at the same time.

\textbf{Interpretability:} The chosen analysis model is a critical component in the pipeline of visual analytics~\cite{keim2008visual}. In the pursuit of better analytical performance, recently developed visual analytics tools tend to leverage advanced machine learning or deep learning models with considerably high complexity. These, however, introduce issues of interpretability of the analysis results and a lack of control over the analytical process, both which are essential for high-impact analytical tasks such as precision medicine and financial investments~\cite{choo2018visual}. To address such problems, there has been an increased research investment towards explainable artificial intelligence~\cite{ming2017understanding,strobelt2017lstmvis}, with the to uncover the inner workings of complex models. Even so, the mechanisms underlying these models can be difficult for non-expert users to understand. Thus, there is a high demand for visual analytics techniques that can organize, transform, and communicate model-level interpretations into comprehensible and actionable guidance. Some recent advancements~\cite{choi2016retain, guo2019visual, carepre} tackle this issue with a focus on a particular analytical tasks and analysis models, yet more generalizable techniques must be explored and developed in future research.

\textbf{Causality Analysis:} From our review of event sequence analysis techniques, we noticed that causality analysis for event sequence data has gained increased attention in the data mining community over the past years. Many causality analysis techniques have been proposed~\cite{xu2016learning, zhang2020cause} to uncover the cause-and-effect relationship between events. However, very few visual analytics techniques have been developed for causality analysis of event sequences. Despite that some existing visual analytics methods are developed for analyzing multivariate data~\cite{wang2015visual, wang2017visual}, the temporal nature and high dimensionality of event sequence data can lead to additional challenges as discussed in Section~\ref{sec:visual_causal}, which is worth addressing in future research. 

\vspace{-3mm}
\section{Conclusion}
\label{sec:conclusion}
This paper presents a survey of visual analytics approaches for event sequence data. The survey proposed a taxonomy that includes a design space and a collection of primary analytical tasks for characterizing the state-of-the-art techniques. In particular, the techniques are partitioned by five categories of analytical tasks, and featured by their corresponding design elements in the design space. It also illustrates the major applications of the techniques through a more domain-specific summary. Finally, the paper discusses the remaining challenges, and points out promising future research directions. With this survey, we connect prior studies in this topic by fitting them together into our taxonomy. We hope our work could provide practitioners with an overview of the alternatives approaches, and help them find the most appropriate design components in developing an effective visual analytics solution that addresses their analytical tasks at hand.

\vspace{-5mm}
\begin{IEEEbiography}[{\includegraphics[width=1in,height=1.25in,clip,keepaspectratio]{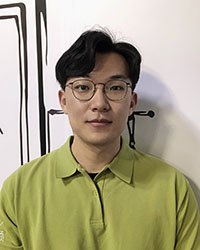}}]{Yi Guo} received his M.S. degree in
Financial Mathematics from the University of 
New South Wales, Australia in 2019. He is currently working
toward his Ph.D. degree as part of the Intelligent Big
Data Visualization (iDV$^{x}$) Lab, Tongji University. His research interests include data visualization and deep learning.
\end{IEEEbiography}
\vspace{-4mm}
\begin{IEEEbiography}[{\includegraphics[width=1in,height=1.25in,clip,keepaspectratio]{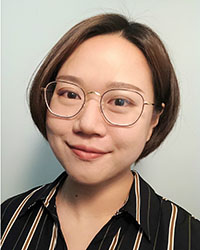}}]{Shunan Guo} received her Ph.D. degree in Software Engineering from East China Normal University, Shanghai, China. Her research interests include visual analytics and human-computer interaction, especially visual analytics approaches for temporal event sequences. For more information, please visit http://guoshunan.com/.
\end{IEEEbiography}
\vspace{-4mm}
\begin{IEEEbiography}[{\includegraphics[width=1in,height=1.25in,clip,keepaspectratio]{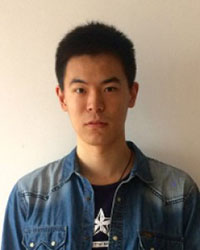}}]{Zhuochen Jin} received his B.S. degree in
Computational Mathematics from Zhejiang
University, China in 2017. He is currently working
toward his Ph.D. degree as part of the Intelligent Big
Data Visualization (iDV$^{x}$) Lab, Tongji University.
His research interests include artificial intelligence and data visualization.
\end{IEEEbiography}
\vspace{-4mm}
\begin{IEEEbiography}[{\includegraphics[width=1in,height=1.25in,clip,keepaspectratio]{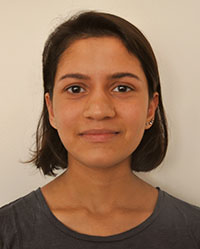}}]{Smiti Kaul} received her Bachelor's degrees in Computer Science and Mathematical Statistics from Wake Forest University, NC, USA. She is currently working towards an M.S. in Computer Science at the University of North Carolina at Chapel Hill, NC, USA, where she is a part of the Visual Analysis and Communication Lab.
\end{IEEEbiography}
\vspace{-4mm}
\begin{IEEEbiography}[{\includegraphics[width=1in,height=1.25in,clip,keepaspectratio]{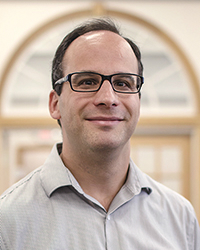}}]{David Gotz} received his Ph.D. in Computer Science from the University of North Carolina (UNC) at Chapel Hill, NC, USA in 2005. He is currently an Associate Professor of Information Science with the School of Information and Library Science at UNC Chapel Hill. He directs the Visual Analysis and Communication Lab and conducts research on a range of topics at the intersection of data visualization, HCI, machine learning, and statistical analysis. He is also the Assistant Director for the Carolina Health Informatics Program and an Associate Member of the UNC Lineberger Comprehensive Cancer Center. He spent nearly a decade as a Research Scientist at the IBM T.J. Watson Research Center, New York, NY, USA before returning to join the UNC faculty in 2014.
\end{IEEEbiography}
\vspace{-5mm}
\begin{IEEEbiography}[{\includegraphics[width=1in,height=1.25in,clip,keepaspectratio]{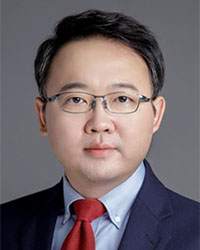}}]{Nan Cao} received his Ph.D. degree in Computer Science and Engineering from the Hong Kong University of Science and Technology (HKUST), Hong Kong, China in 2012. He is currently a professor at Tongji University and the Assistant Dean of the Tongji College of Design and Innovation. He also directs the Tongji Intelligent Big Data Visualization Lab (iDV$^x$ Lab) and conducts interdisciplinary research across multiple fields, including data visualization, human computer interaction, machine learning, and data mining. Before his Ph.D. studies at HKUST, he was a staff researcher at IBM China Research Lab, Beijing, China. He was a research staff member at the IBM T.J. Watson Research Center, New York, NY, USA before joining the Tongji faculty in 2016.
\end{IEEEbiography}
\begin{spacing}{1}
\bibliographystyle{abbrv-doi}

\begin{thebibliography}{100}

\bibitem{bhattacharjya2020event}
D.~Bhattacharjya, K.~Shanmugam, T.~Gao, N.~Mattei, K.~R. Varshney, and
  D.~Subramanian.
\newblock Event-driven continuous time bayesian networks.
\newblock In {\em AAAI Conference on Artificial Intelligence}, pp. 3259--3266,
  2020.

\bibitem{brehmer2016timelines}
M.~Brehmer, B.~Lee, B.~Bach, N.~H. Riche, and T.~Munzner.
\newblock Timelines revisited: A design space and considerations for expressive
  storytelling.
\newblock {\em IEEE Transactions on Visualization and Computer Graphics},
  23(9):2151--2164, 2016.

\bibitem{cao2011dicon}
N.~Cao, D.~Gotz, J.~Sun, and H.~Qu.
\newblock Dicon: Interactive visual analysis of multidimensional clusters.
\newblock {\em IEEE Transactions on Visualization and Computer Graphics},
  17(12):2581--2590, 2011.

\bibitem{cao2015episogram}
N.~Cao, Y.-R. Lin, F.~Du, and D.~Wang.
\newblock Episogram: Visual summarization of egocentric social interactions.
\newblock {\em IEEE Computer Graphics and Applications}, 36(5):72--81, 2015.

\bibitem{cao2012whisper}
N.~Cao, Y.-R. Lin, X.~Sun, D.~Lazer, S.~Liu, and H.~Qu.
\newblock Whisper: Tracing the spatiotemporal process of information diffusion
  in real time.
\newblock {\em IEEE Transactions on Visualization and Computer Graphics},
  18(12):2649--2658, 2012.

\bibitem{cao2015targetvue}
N.~Cao, C.~Shi, S.~Lin, J.~Lu, Y.-R. Lin, and C.-Y. Lin.
\newblock Targetvue: Visual analysis of anomalous user behaviors in online
  communication systems.
\newblock {\em IEEE Transactions on Visualization and Computer Graphics},
  22(1):280--289, 2015.

\bibitem{cappers2017exploring}
B.~C. Cappers and J.~J. van Wijk.
\newblock Exploring multivariate event sequences using rules, aggregations, and
  selections.
\newblock {\em IEEE Transactions on Visualization and Computer Graphics},
  24(1):532--541, 2017.

\bibitem{chankhihort2017visualization}
D.~Chankhihort, B.-M. Lim, G.-J. Lee, S.~Choi, S.-O. Kwon, S.-H. Lee, J.-T.
  Kang, A.~Nasridinov, and K.-H. Yoo.
\newblock A visualization scheme with a calendar heat map for abnormal pattern
  analysis in the manufacturing process.
\newblock {\em International Journal of Contents}, 13(2):21--28, 2017.

\bibitem{chen2011data}
M.~Chen, A.~Trefethen, R.~Banares-Alcantara, M.~Jirotka, B.~Coecke, T.~Ertl,
  and A.~Schmidt.
\newblock From data analysis and visualization to causality discovery.
\newblock {\em Computer}, (10):84--87, 2011.

\bibitem{chen2015peakvizor}
Q.~Chen, Y.~Chen, D.~Liu, C.~Shi, Y.~Wu, and H.~Qu.
\newblock Peakvizor: Visual analytics of peaks in video clickstreams from
  massive open online courses.
\newblock {\em IEEE Transactions on Visualization and Computer Graphics},
  22(10):2315--2330, 2015.

\bibitem{chen2018d}
S.~Chen, S.~Chen, Z.~Wang, J.~Liang, Y.~Wu, and X.~Yuan.
\newblock D-map+ interactive visual analysis and exploration of ego-centric and
  event-centric information diffusion patterns in social media.
\newblock {\em ACM Transactions on Intelligent Systems and Technology},
  10(1):1--26, 2018.

\bibitem{chen2019r}
S.~Chen, S.~Li, S.~Chen, and X.~Yuan.
\newblock R-map: A map metaphor for visualizing information reposting process
  in social media.
\newblock {\em IEEE Transactions on Visualization and Computer Graphics},
  26(1):1204--1214, 2019.

\bibitem{chen2018stagemap}
Y.~Chen, A.~Puri, L.~Yuan, and H.~Qu.
\newblock Stagemap: Extracting and summarizing progression stages in event
  sequences.
\newblock In {\em IEEE International Conference on Big Data}, pp. 975--981.
  IEEE, 2018.

\bibitem{chen2018sequence}
Y.~Chen, P.~Xu, and L.~Ren.
\newblock Sequence synopsis: Optimize visual summary of temporal event data.
\newblock {\em IEEE Transactions on Visualization and Computer Graphics},
  24(1):45--55, 2017.

\bibitem{choi2016retain}
E.~Choi, M.~T. Bahadori, J.~Sun, J.~Kulas, A.~Schuetz, and W.~Stewart.
\newblock Retain: An interpretable predictive model for healthcare using
  reverse time attention mechanism.
\newblock In {\em Advances in Neural Information Processing Systems}, pp.
  3504--3512. Curran Associates, 2016.

\bibitem{choo2018visual}
J.~Choo and S.~Liu.
\newblock Visual analytics for explainable deep learning.
\newblock {\em IEEE Computer Graphics and Applications}, 38(4):84--92, 2018.

\bibitem{cook2005illuminating}
K.~A. Cook and J.~J. Thomas.
\newblock Illuminating the path: The research and development agenda for visual
  analytics.
\newblock Technical report, Pacific Northwest National Lab.(PNNL), Richland, WA
  (United States), 2005.

\bibitem{dang2015reactionflow}
T.~N. Dang, P.~Murray, J.~Aurisano, and A.~G. Forbes.
\newblock Reactionflow: an interactive visualization tool for causality
  analysis in biological pathways.
\newblock In {\em BMC proceedings}, vol.~9, p.~S6. BioMed Central, 2015.

\bibitem{du2020interactive}
F.~Du, S.~Guo, S.~Malik, E.~Koh, S.~Kim, and Z.~Liu.
\newblock Interactive event sequence prediction for marketing analysts.
\newblock In {\em Extended Abstracts of the CHI Conference on Human Factors in
  Computing Systems}, pp. 1--8, 2020.

\bibitem{du2016eventaction}
F.~Du, C.~Plaisant, N.~Spring, and B.~Shneiderman.
\newblock Eventaction: Visual analytics for temporal event sequence
  recommendation.
\newblock In {\em Visual Analytics Science and Technology}, pp. 61--70. IEEE,
  2016.

\bibitem{du2017finding}
F.~Du, C.~Plaisant, N.~Spring, and B.~Shneiderman.
\newblock Finding similar people to guide life choices: Challenge, design, and
  evaluation.
\newblock In {\em Proceedings of the CHI Conference on Human Factors in
  Computing Systems}, pp. 5498--5544. ACM, 2017.

\bibitem{du2018visual}
F.~Du, C.~Plaisant, N.~Spring, and B.~Shneiderman.
\newblock Visual interfaces for recommendation systems: Finding similar and
  dissimilar peers.
\newblock {\em ACM Transactions on Intelligent Systems and Technology},
  10(1):1--23, 2018.

\bibitem{elmqvist2003growing}
N.~Elmqvist and P.~Tsigas.
\newblock Growing squares: Animated visualization of causal relations.
\newblock In {\em Proceedings of the ACM symposium on Software Visualization},
  pp. 17--ff, 2003.

\bibitem{elmqvist2004animated}
N.~Elmqvist and P.~Tsigas.
\newblock Animated visualization of causal relations through growing 2d
  geometry.
\newblock {\em Information Visualization}, 3(3):154--172, 2004.

\bibitem{fails2006visual}
J.~A. Fails, A.~Karlson, L.~Shahamat, and B.~Shneiderman.
\newblock A visual interface for multivariate temporal data: Finding patterns
  of events across multiple histories.
\newblock In {\em IEEE Symposium On Visual Analytics Science And Technology},
  pp. 167--174. IEEE, 2006.

\bibitem{fan2017personal}
X.~Fan, Y.~Peng, Y.~Zhao, Y.~Li, D.~Meng, Z.~Zhong, F.~Zhou, and M.~Lu.
\newblock A personal visual analytics on smartphone usage data.
\newblock {\em Journal of Visual Languages \& Computing}, 41:111--120, 2017.

\bibitem{fischer2012vistracer}
F.~Fischer, J.~Fuchs, P.-A. Vervier, F.~Mansmann, and O.~Thonnard.
\newblock Vistracer: a visual analytics tool to investigate routing anomalies
  in traceroutes.
\newblock In {\em Proceedings of the International Symposium on Visualization
  for Cyber Security}, pp. 80--87, 2012.

\bibitem{franklin2016treatmentexplorer}
L.~Franklin, C.~Plaisant, K.~Minhazur~Rahman, and B.~Shneiderman.
\newblock Treatmentexplorer: An interactive decision aid for medical risk
  communication and treatment exploration.
\newblock {\em Interacting with Computers}, 28(3):238--252, 2016.

\bibitem{gotz2014decisionflow}
D.~Gotz and H.~Stavropoulos.
\newblock Decisionflow: Visual analytics for high-dimensional temporal event
  sequence data.
\newblock {\em IEEE Transactions on Visualization and Computer Graphics},
  20(12):1783--1792, 2014.

\bibitem{gotz2016adaptive}
D.~Gotz, S.~Sun, and N.~Cao.
\newblock Adaptive contextualization: Combating bias during high-dimensional
  visualization and data selection.
\newblock In {\em Proceedings of the 21st International Conference on
  Intelligent User Interfaces}, pp. 85--95. ACM, 2016.

\bibitem{gotz2020}
D.~Gotz, J.~Zhang, W.~Wang, and J.~Shrestha.
\newblock Visual analysis of high-dimensional event sequence data via dynamic
  hierarchical aggregation.
\newblock {\em IEEE Transactions on Visualization and Computer Graphics},
  26(1), 2020.

\bibitem{goulden2019ccvis}
M.~C. Goulden, E.~Gronda, Y.~Yang, Z.~Zhang, J.~Tao, C.~Wang, X.~Duan, G.~A.
  Ambrose, K.~Abbott, and P.~Miller.
\newblock Ccvis: Visual analytics of student online learning behaviors using
  course clickstream data.
\newblock {\em Electronic Imaging}, 2019(1):681--1, 2019.

\bibitem{guo2020comparative}
R.~Guo, T.~Fujiwara, Y.~Li, K.~M. Lima, S.~Sen, N.~K. Tran, and K.-L. Ma.
\newblock Comparative visual analytics for assessing medical records with
  sequence embedding.
\newblock {\em Visual Informatics}, 2020.

\bibitem{guo2019visualizing}
S.~Guo, F.~Du, S.~Malik, E.~Koh, S.~Kim, Z.~Liu, D.~Kim, H.~Zha, and N.~Cao.
\newblock Visualizing uncertainty and alternatives in event sequence
  predictions.
\newblock In {\em Proceedings of the CHI Conference on Human Factors in
  Computing Systems}, pp. 1--12, 2019.

\bibitem{guo2019visual}
S.~Guo, Z.~Jin, Q.~Chen, D.~Gotz, H.~Zha, and N.~Cao.
\newblock Visual anomaly detection in event sequence data.
\newblock In {\em IEEE International Conference on Big Data}, pp. 1125--1130.
  IEEE, 2019.

\bibitem{guo2018visual}
S.~Guo, Z.~Jin, D.~Gotz, F.~Du, H.~Zha, and N.~Cao.
\newblock Visual progression analysis of event sequence data.
\newblock {\em IEEE Transactions on Visualization and Computer Graphics}, pp.
  1--1, 2018.

\bibitem{guo2017eventthread}
S.~Guo, K.~Xu, R.~Zhao, D.~Gotz, H.~Zha, and N.~Cao.
\newblock Eventthread: Visual summarization and stage analysis of event
  sequence data.
\newblock {\em IEEE Transactions on Visualization and Computer Graphics},
  24(1):56--65, 2017.

\bibitem{han2015visual}
Y.~Han, A.~Rozga, N.~Dimitrova, G.~D. Abowd, and J.~Stasko.
\newblock Visual analysis of proximal temporal relationships of social and
  communicative behaviors.
\newblock In {\em Computer Graphics Forum}, vol.~34, pp. 51--60. Wiley Online
  Library, 2015.

\bibitem{he2019vuc}
H.~He, B.~Dong, Q.~Zheng, and G.~Li.
\newblock Vuc: Visualizing daily video utilization to promote student
  engagement in online distance education.
\newblock In {\em Proceedings of the ACM Conference on Global Computing
  Education}, pp. 99--105, 2019.

\bibitem{herr2018visual}
D.~Herr, F.~Beck, and T.~Ertl.
\newblock Visual analytics for decomposing temporal event series of production
  lines.
\newblock In {\em 22nd International Conference Information Visualisation}, pp.
  251--259. IEEE, 2018.

\bibitem{hibbs}
M.~A. Hibbs, N.~C. Dirksen, K.~Li, and O.~G. Troyanskaya.
\newblock Visualization methods for statistical analysis of microarray
  clusters.
\newblock {\em BMC Bioinformatics}, 6(1):115, 2005.

\bibitem{huisman2005complexities}
O.~Huisman and P.~Forer.
\newblock The complexities of everyday life: balancing practical and realistic
  approaches to modeling probable presence in space-time.
\newblock In {\em The Annual Colloquium of the Spatial Information Research
  Centre}, pp. 155--167. Citeseer, 2005.

\bibitem{jentner2019visualization}
W.~Jentner and D.~A. Keim.
\newblock Visualization and visual analytic techniques for patterns.
\newblock In {\em High-Utility Pattern Mining}, pp. 303--337. Springer, 2019.

\bibitem{carepre}
Z.~Jin, S.~Cui, S.~Guo, D.~Gotz, J.~Sun, and N.~Cao.
\newblock Carepre: An intelligent clinical decision assistance system.
\newblock {\em ACM Transactions on Computing for Healthcare}, 1(1):1--20, 2020.

\bibitem{jo2014livegantt}
J.~Jo, J.~Huh, J.~Park, B.~Kim, and J.~Seo.
\newblock Livegantt: Interactively visualizing a large manufacturing schedule.
\newblock {\em IEEE Transactions on Visualization and Computer Graphics},
  20(12):2329--2338, 2014.

\bibitem{kadaba2007visualizing}
N.~R. Kadaba, P.~P. Irani, and J.~Leboe.
\newblock Visualizing causal semantics using animations.
\newblock {\em IEEE Transactions on Visualization and Computer Graphics},
  13(6):1254--1261, 2007.

\bibitem{kandel2011research}
S.~Kandel, J.~Heer, C.~Plaisant, J.~Kennedy, F.~Van~Ham, N.~H. Riche,
  C.~Weaver, B.~Lee, D.~Brodbeck, and P.~Buono.
\newblock Research directions in data wrangling: Visualizations and
  transformations for usable and credible data.
\newblock {\em Information Visualization}, 10(4):271--288, 2011.

\bibitem{keim2008visual}
D.~Keim, G.~Andrienko, J.-D. Fekete, C.~G{\"o}rg, J.~Kohlhammer, and
  G.~Melan{\c{c}}on.
\newblock Visual analytics: Definition, process, and challenges.
\newblock In {\em Information Visualization}, pp. 154--175. Springer, 2008.

\bibitem{kiernan2009constructing}
J.~Kiernan and E.~Terzi.
\newblock Constructing comprehensive summaries of large event sequences.
\newblock {\em ACM Transactions on Knowledge Discovery from Data}, 3(4):21,
  2009.

\bibitem{koldehofe1999distributed}
B.~Koldehofe, M.~Papatriantafilou, and P.~Tsigas.
\newblock Distributed algorithms visualisation for educational purposes.
\newblock In {\em Proceedings of the Annual SIGCSE/SIGCUE ITiCSE Conference on
  Innovation and Technology in Computer Science Education}, pp. 103--106, 1999.

\bibitem{krause2016interacting}
J.~Krause, A.~Perer, and K.~Ng.
\newblock Interacting with predictions: Visual inspection of black-box machine
  learning models.
\newblock In {\em Proceedings of the CHI Conference on Human Factors in
  Computing Systems}, pp. 5686--5697. ACM, New York, 2016.

\bibitem{krause2015supporting}
J.~Krause, A.~Perer, and H.~Stavropoulos.
\newblock Supporting iterative cohort construction with visual temporal
  queries.
\newblock {\em IEEE Transactions on Visualization and Computer Graphics},
  22(1):91--100, 2015.

\bibitem{kwan1}
M.-P. Kwan.
\newblock Gender and individual access to urban opportunities: a study using
  space--time measures.
\newblock {\em The Professional Geographer}, 51(2):210--227, 1999.

\bibitem{kwon2020dpvis}
B.~C. Kwon, V.~Anand, K.~A. Severson, S.~Ghosh, Z.~Sun, B.~I. Frohnert,
  M.~Lundgren, and K.~Ng.
\newblock Dpvis: Visual analytics with hidden markov models for disease
  progression pathways.
\newblock {\em IEEE Transactions on Visualization and Computer Graphics}, 2020.

\bibitem{kwon2018retainvis}
B.~C. Kwon, M.-J. Choi, J.~T. Kim, E.~Choi, Y.~B. Kim, S.~Kwon, J.~Sun, and
  J.~Choo.
\newblock Retainvis: Visual analytics with interpretable and interactive
  recurrent neural networks on electronic medical records.
\newblock {\em IEEE Transactions on Visualization and Computer Graphics}, 2018.

\bibitem{kwon2016peekquence}
B.~C. Kwon, J.~Verma, and A.~Perer.
\newblock Peekquence: Visual analytics for event sequence data.
\newblock In {\em ACM SIGKDD 2016 Workshop on Interactive Data Exploration and
  Analytics}, vol.~1, 2016.

\bibitem{law2018maqui}
P.-M. Law, Z.~Liu, S.~Malik, and R.~C. Basole.
\newblock Maqui: Interweaving queries and pattern mining for recursive event
  sequence exploration.
\newblock {\em IEEE Transactions on Visualization and Computer Graphics},
  25(1):396--406, 2018.

\bibitem{li2019visualizing}
W.~Li, M.~Funk, Q.~Li, and A.~Brombacher.
\newblock Visualizing event sequence game data to understand player’s skill
  growth through behavior complexity.
\newblock {\em Journal of Visualization}, 22(4):833--850, 2019.

\bibitem{liu2016mining}
Z.~Liu, H.~Dev, M.~Dontcheva, and M.~Hoffman.
\newblock Mining, pruning and visualizing frequent patterns for temporal event
  sequence analysis.
\newblock In {\em Proceedings of the IEEE VIS Workshop on Temporal \&
  Sequential Event Analysis}, pp. 2--4, 2016.

\bibitem{liu2017coreflow}
Z.~Liu, B.~Kerr, M.~Dontcheva, J.~Grover, M.~Hoffman, and A.~Wilson.
\newblock Coreflow: Extracting and visualizing branching patterns from event
  sequences.
\newblock In {\em Computer Graphics Forum}, vol.~36, pp. 527--538. Wiley Online
  Library, 2017.

\bibitem{liu2016patterns}
Z.~Liu, Y.~Wang, M.~Dontcheva, M.~Hoffman, S.~Walker, and A.~Wilson.
\newblock Patterns and sequences: Interactive exploration of clickstreams to
  understand common visitor paths.
\newblock {\em IEEE Transactions on Visualization and Computer Graphics},
  23(1):321--330, 2016.

\bibitem{malik2015cohort}
S.~Malik, F.~Du, M.~Monroe, E.~Onukwugha, C.~Plaisant, and B.~Shneiderman.
\newblock Cohort comparison of event sequences with balanced integration of
  visual analytics and statistics.
\newblock In {\em Proceedings of the 20th International Conference on
  Intelligent User Interfaces}, pp. 38--49, 2015.

\bibitem{malik2016high}
S.~Malik, B.~Shneiderman, F.~Du, C.~Plaisant, and M.~Bjarnadottir.
\newblock High-volume hypothesis testing: Systematic exploration of event
  sequence comparisons.
\newblock {\em ACM Transactions on Interactive Intelligent Systems},
  6(1):1--23, 2016.

\bibitem{mei2017neural}
H.~Mei and J.~M. Eisner.
\newblock The neural hawkes process: A neurally self-modulating multivariate
  point process.
\newblock In {\em Advances in Neural Information Processing Systems}, pp.
  6754--6764, 2017.

\bibitem{michotte1963causalite}
A.~Michotte, G.~Thines, A.~Costall, and G.~Butterworth.
\newblock La causalit{\'e} perceptive.
\newblock {\em Journal de Psychologie Normale Et Pathologique}, 60:9--36, 1963.

\bibitem{ming2017understanding}
Y.~Ming, S.~Cao, R.~Zhang, Z.~Li, Y.~Chen, Y.~Song, and H.~Qu.
\newblock Understanding hidden memories of recurrent neural networks.
\newblock In {\em IEEE Conference on Visual Analytics Science and Technology},
  pp. 13--24. IEEE, 2017.

\bibitem{monroe2014interactive}
M.~Monroe.
\newblock {\em Interactive event sequence query and transformation}.
\newblock PhD thesis, 2014.

\bibitem{mu2019moocad}
X.~Mu, K.~Xu, Q.~Chen, F.~Du, Y.~Wang, and H.~Qu.
\newblock Moocad: Visual analysis of anomalous learning activities in massive
  open online courses.
\newblock In {\em Eurographics/IEEE VGTC Conference on Visualization}, pp.
  91--95, 2019.

\bibitem{muelder2016visual}
C.~Muelder, B.~Zhu, W.~Chen, H.~Zhang, and K.-L. Ma.
\newblock Visual analysis of cloud computing performance using behavioral
  lines.
\newblock {\em IEEE Transactions on Visualization and Computer Graphics},
  22(6):1694--1704, 2016.

\bibitem{ng2014paramo}
K.~Ng, A.~Ghoting, S.~R. Steinhubl, W.~F. Stewart, B.~Malin, and J.~Sun.
\newblock Paramo: A parallel predictive modeling platform for healthcare
  analytic research using electronic health records.
\newblock {\em Journal of biomedical informatics}, 48:160--170, 2014.

\bibitem{vasabi}
P.~H. Nguyen, R.~Henkin, S.~Chen, N.~Andrienko, G.~Andrienko, O.~Thonnard, and
  C.~Turkay.
\newblock Vasabi: Hierarchical user profiles for interactive visual user
  behaviour analytics.
\newblock {\em IEEE Transactions on Visualization and Computer Graphics},
  26(1), 2020.

\bibitem{nguyen2018understanding}
P.~H. Nguyen, C.~Turkay, G.~Andrienko, N.~Andrienko, O.~Thonnard, and
  J.~Zouaoui.
\newblock Understanding user behaviour through action sequences: from the usual
  to the unusual.
\newblock {\em IEEE Transactions on Visualization and Computer Graphics},
  25(9):2838--2852, 2018.

\bibitem{nielsen2009abyss}
C.~B. Nielsen, S.~D. Jackman, I.~Birol, and S.~J. Jones.
\newblock Abyss-explorer: visualizing genome sequence assemblies.
\newblock {\em IEEE Transactions on Visualization and Computer Graphics},
  15(6):881--888, 2009.

\bibitem{perer2013data}
A.~Perer and D.~Gotz.
\newblock Data-driven exploration of care plans for patients.
\newblock In {\em Extended Abstracts of the CHI Conference on Human Factors in
  Computing Systems}, pp. 439--444. ACM, Paris, 2013.

\bibitem{perer2012matrixflow}
A.~Perer and J.~Sun.
\newblock Matrixflow: temporal network visual analytics to track symptom
  evolution during disease progression.
\newblock In {\em AMIA Annual Symposium Proceedings}, vol. 2012, p. 716.
  American Medical Informatics Association, 2012.

\bibitem{perer2014frequence}
A.~Perer and F.~Wang.
\newblock Frequence: Interactive mining and visualization of temporal frequent
  event sequences.
\newblock In {\em Proceedings of the International Conference on Intelligent
  User Interfaces}, pp. 153--162. ACM, 2014.

\bibitem{perer2015mining}
A.~Perer, F.~Wang, and J.~Hu.
\newblock Mining and exploring care pathways from electronic medical records
  with visual analytics.
\newblock {\em Journal of Biomedical Informatics}, 56:369--378, 2015.

\bibitem{plaisant1996lifelines}
C.~Plaisant, B.~Milash, A.~Rose, S.~Widoff, and B.~Shneiderman.
\newblock Lifelines: visualizing personal histories.
\newblock In {\em Proceedings of the CHI Conference on Human Factors in
  Computing Systems}, pp. 221--227. ACM, 1996.

\bibitem{polack2015timestitch}
P.~J. Polack, S.-T. Chen, M.~Kahng, M.~Sharmin, and D.~H. Chau.
\newblock Timestitch: Interactive multi-focus cohort discovery and comparison.
\newblock In {\em IEEE Conference on Visual Analytics Science and Technology},
  pp. 209--210, 2015.

\bibitem{polack2018chronodes}
P.~J. Polack~Jr, S.-T. Chen, M.~Kahng, K.~D. Barbaro, R.~Basole, M.~Sharmin,
  and D.~H. Chau.
\newblock Chronodes: Interactive multifocus exploration of event sequences.
\newblock {\em ACM Transactions on Interactive Intelligent Systems},
  8(1):1--21, 2018.

\bibitem{qi2019stbins}
J.~Qi, V.~Bloemen, S.~Wang, J.~van Wijk, and H.~van~de Wetering.
\newblock Stbins: visual tracking and comparison of multiple data sequences
  using temporal binning.
\newblock {\em IEEE Transactions on visualization and computer graphics},
  26(1):1054--1063, 2019.

\bibitem{riehmann2005interactive}
P.~Riehmann, M.~Hanfler, and B.~Froehlich.
\newblock Interactive sankey diagrams.
\newblock In {\em IEEE Symposium on Information Visualization}, pp. 233--240.
  IEEE, 2005.

\bibitem{robinson}
A.~C. Robinson, D.~J. Peuquet, S.~Pezanowski, F.~A. Hardisty, and B.~Swedberg.
\newblock Design and evaluation of a geovisual analytics system for uncovering
  patterns in spatio-temporal event data.
\newblock {\em Cartography and Geographic Information Science}, 44(3):216--228,
  2017.

\bibitem{rogers2019composer}
J.~Rogers, N.~Spina, A.~Neese, R.~Hess, D.~Brodke, and A.~Lex.
\newblock Composer—visual cohort analysis of patient outcomes.
\newblock {\em Applied Clinical Informatics}, 10(02):278--285, 2019.

\bibitem{Rosenthal2013}
P.~Rosenthal, L.~Pfeiffer, N.~H. Müller, and P.~Ohler.
\newblock Visruption: Intuitive and efficient visualization of temporal airline
  disruption data.
\newblock {\em Computer Graphics Forum}, 2013.

\bibitem{rzes}
J.~Rzeszotarski and A.~Kittur.
\newblock Crowdscape: Interactively visualizing user behavior and output.
\newblock In {\em Proceedings of the Annual ACM Symposium on User Interface
  Software and Technology}, p. 55–62, 2012.

\bibitem{sacha2015role}
D.~Sacha, H.~Senaratne, B.~C. Kwon, G.~Ellis, and D.~A. Keim.
\newblock The role of uncertainty, awareness, and trust in visual analytics.
\newblock {\em IEEE Transactions on Visualization and Computer Graphics},
  22(1):240--249, 2015.

\bibitem{saraiya}
P.~{Saraiya}, C.~{North}, and K.~{Duca}.
\newblock An evaluation of microarray visualization tools for biological
  insight.
\newblock In {\em IEEE Symposium on Information Visualization}, pp. 1--8, 2004.

\bibitem{Sarikaya2016}
A.~Sarikaya, M.~Correll, J.~M. Dinis, D.~H. O'Connor, and M.~Gleicher.
\newblock {Visualizing Co-occurrence of Events in Populations of Viral Genome
  Sequences}.
\newblock {\em Computer Graphics Forum}, 2016.

\bibitem{senin2015time}
P.~Senin, J.~Lin, X.~Wang, T.~Oates, S.~Gandhi, A.~P. Boedihardjo, C.~Chen, and
  S.~Frankenstein.
\newblock Time series anomaly discovery with grammar-based compression.
\newblock In {\em International Conference on Extending Database Technology},
  pp. 481--492, 2015.

\bibitem{seo2002interactively}
J.~Seo and B.~Shneiderman.
\newblock Interactively exploring hierarchical clustering results [gene
  identification].
\newblock {\em Computer}, 35(7):80--86, 2002.

\bibitem{shi2019visual}
Y.~Shi, Y.~Liu, H.~Tong, J.~He, G.~Yan, and N.~Cao.
\newblock Visual analytics of anomalous user behaviors: A survey.
\newblock {\em IEEE Transactions on Big Data}, pp. 1--1, 2020.

\bibitem{shneiderman1996eyes}
B.~Shneiderman.
\newblock The eyes have it: A task by data type taxonomy for information
  visualizations.
\newblock In {\em Proceedings IEEE Symposium on Visual Languages}, pp.
  336--343, 1996.

\bibitem{slack2004sequencejuxtaposer}
J.~Slack, K.~Hildebrand, T.~Munzner, and K.~S. John.
\newblock Sequencejuxtaposer: Fluid navigation for large-scale sequence
  comparison in context.
\newblock In {\em German Conference on Bioinformatics}, pp. 37--42, 2004.

\bibitem{stopar2018streamstory}
L.~Stopar, P.~Skraba, M.~Grobelnik, and D.~Mladenic.
\newblock Streamstory: exploring multivariate time series on multiple scales.
\newblock {\em IEEE Transactions on Visualization and Computer Graphics},
  25(4):1788--1802, 2018.

\bibitem{strobelt2017lstmvis}
H.~Strobelt, S.~Gehrmann, H.~Pfister, and A.~M. Rush.
\newblock Lstmvis: A tool for visual analysis of hidden state dynamics in
  recurrent neural networks.
\newblock {\em IEEE Transactions on Visualization and Computer Graphics},
  24(1):667--676, 2017.

\bibitem{sun2019planningvis}
D.~Sun, R.~Huang, Y.~Chen, Y.~Wang, J.~Zeng, M.~Yuan, T.-C. Pong, and H.~Qu.
\newblock Planningvis: A visual analytics approach to production planning in
  smart factories.
\newblock {\em IEEE Transactions on Visualization and Computer Graphics}, 2019.

\bibitem{sun2017socialwave}
G.~Sun, T.~Tang, T.-Q. Peng, R.~Liang, and Y.~Wu.
\newblock Socialwave: visual analysis of spatio-temporal diffusion of
  information on social media.
\newblock {\em ACM Transactions on Intelligent Systems and Technology},
  9(2):1--23, 2017.

\bibitem{sun2013survey}
G.-D. Sun, Y.-C. Wu, R.-H. Liang, and S.-X. Liu.
\newblock A survey of visual analytics techniques and applications:
  State-of-the-art research and future challenges.
\newblock {\em Journal of Computer Science and Technology}, 28(5):852--867,
  2013.

\bibitem{trumper2012viewfusion}
J.~Tr{\"u}mper, A.~Telea, and J.~D{\"o}llner.
\newblock Viewfusion: Correlating structure and activity views for execution
  traces.
\newblock In {\em Theory and Practice of Computer Graphics}, pp. 45--52. The
  Eurographics Association, 2012.

\bibitem{viegas2013google}
F.~Vi{\'e}gas, M.~Wattenberg, J.~Hebert, G.~Borggaard, A.~Cichowlas,
  J.~Feinberg, J.~Orwant, and C.~Wren.
\newblock Google+ ripples: A native visualization of information flow.
\newblock In {\em Proceedings of the 22nd international conference on World
  Wide Web}, pp. 1389--1398, 2013.

\bibitem{vrotsou}
K.~Vrotsou.
\newblock {\em Everyday mining: Exploring sequences in event-based data}.
\newblock Doctoral thesis, Linkoping University, The Institute of Technology,
  2010.

\bibitem{vrotsou2009activitree}
K.~Vrotsou, J.~Johansson, and M.~Cooper.
\newblock Activitree: Interactive visual exploration of sequences in
  event-based data using graph similarity.
\newblock {\em IEEE Transactions on Visualization and Computer Graphics},
  15(6):945--952, 2009.

\bibitem{wang}
G.~Wang, X.~Zhang, S.~Tang, H.~Zheng, and B.~Y. Zhao.
\newblock Unsupervised clickstream clustering for user behavior analytics.
\newblock In {\em Proceedings of the CHI Conference on Human Factors in
  Computing Systems}, pp. 225--236, 2016.

\bibitem{wang2015visual}
J.~Wang and K.~Mueller.
\newblock The visual causality analyst: An interactive interface for causal
  reasoning.
\newblock {\em IEEE Transactions on Visualization and Computer Graphics},
  22(1):230--239, 2015.

\bibitem{wang2017visual}
J.~Wang and K.~Mueller.
\newblock Visual causality analysis made practical.
\newblock In {\em IEEE Conference on Visual Analytics Science and Technology},
  pp. 151--161, 2017.

\bibitem{wang2008aligning}
T.~D. Wang, C.~Plaisant, A.~J. Quinn, R.~Stanchak, S.~Murphy, and
  B.~Shneiderman.
\newblock Aligning temporal data by sentinel events: discovering patterns in
  electronic health records.
\newblock In {\em Proceedings of the CHI Conference on Human Factors in
  Computing Systems}, pp. 457--466, 2008.

\bibitem{wang2009temporal}
T.~D. Wang, C.~Plaisant, B.~Shneiderman, N.~Spring, D.~Roseman, G.~Marchand,
  V.~Mukherjee, and M.~Smith.
\newblock Temporal summaries: Supporting temporal categorical searching,
  aggregation and comparison.
\newblock {\em IEEE Transactions on Visualization and Computer Graphics},
  15(6):1049--1056, 2009.

\bibitem{wang2009visualization}
X.~Wang, W.~Dou, W.~Ribarsky, and R.~Chang.
\newblock Visualization as integration of heterogeneous processes.
\newblock In {\em Visual Analytics for Homeland Defense and Security}, vol.
  7346, p. 73460B. International Society for Optics and Photonics, 2009.

\bibitem{wei2012visual}
J.~Wei, Z.~Shen, N.~Sundaresan, and K.-L. Ma.
\newblock Visual cluster exploration of web clickstream data.
\newblock In {\em IEEE Conference on Visual Analytics Science and Technology},
  pp. 3--12, 2012.

\bibitem{wongsuphasawat2011outflow}
K.~Wongsuphasawat and D.~Gotz.
\newblock Outflow: Visualizing patient flow by symptoms and outcome.
\newblock In {\em IEEE VisWeek Workshop on Visual Analytics in Healthcare}, pp.
  25--28. American Medical Informatics Association, 2011.

\bibitem{wongsuphasawat2012exploring}
K.~Wongsuphasawat and D.~Gotz.
\newblock Exploring flow, factors, and outcomes of temporal event sequences
  with the outflow visualization.
\newblock {\em IEEE Transactions on Visualization and Computer Graphics},
  18(12):2659--2668, 2012.

\bibitem{wongsuphasawat2011lifeflow}
K.~Wongsuphasawat, J.~A. Guerra~G{\'o}mez, C.~Plaisant, T.~D. Wang,
  M.~Taieb-Maimon, and B.~Shneiderman.
\newblock Lifeflow: visualizing an overview of event sequences.
\newblock In {\em Proceedings of the CHI Conference on Human Factors in
  Computing Systems}, pp. 1747--1756, 2011.

\bibitem{wongsuphasawat2009finding}
K.~Wongsuphasawat and B.~Shneiderman.
\newblock Finding comparable temporal categorical records: A similarity measure
  with an interactive visualization.
\newblock In {\em 2009 IEEE Symposium on Visual Analytics Science and
  Technology}, pp. 27--34, 2009.

\bibitem{wu2018visual}
W.~Wu, Y.~Zheng, K.~Chen, X.~Wang, and N.~Cao.
\newblock A visual analytics approach for equipment condition monitoring in
  smart factories of process industry.
\newblock In {\em IEEE Pacific Visualization Proceedings}, pp. 140--149, 2018.

\bibitem{wu2016survey}
Y.~Wu, N.~Cao, D.~Gotz, Y.-P. Tan, and D.~A. Keim.
\newblock A survey on visual analytics of social media data.
\newblock {\em IEEE Transactions on Multimedia}, 18(11):2135--2148, 2016.

\bibitem{wu2014opinionflow}
Y.~Wu, S.~Liu, K.~Yan, M.~Liu, and F.~Wu.
\newblock Opinionflow: Visual analysis of opinion diffusion on social media.
\newblock {\em IEEE Transactions on Visualization and Computer Graphics},
  20(12):1763--1772, 2014.

\bibitem{xie2018visual}
C.~Xie, W.~Xu, and K.~Mueller.
\newblock A visual analytics framework for the detection of anomalous call
  stack trees in high performance computing applications.
\newblock {\em IEEE Transactions on Visualization and Computer Graphics},
  25(1):215--224, 2018.

\bibitem{xu2016learning}
H.~Xu, M.~Farajtabar, and H.~Zha.
\newblock Learning granger causality for hawkes processes.
\newblock In {\em International Conference on Machine Learning}, pp.
  1717--1726, 2016.

\bibitem{xu2019clouddet}
K.~Xu, Y.~Wang, L.~Yang, Y.~Wang, B.~Qiao, S.~Qin, Y.~Xu, H.~Zhang, and H.~Qu.
\newblock Clouddet: Interactive visual analysis of anomalous performances in
  cloud computing systems.
\newblock {\em IEEE Transactions on Visualization and Computer Graphics},
  26(1):1107--1117, 2019.

\bibitem{xu2016vidx}
P.~Xu, H.~Mei, L.~Ren, and W.~Chen.
\newblock Vidx: Visual diagnostics of assembly line performance in smart
  factories.
\newblock {\em IEEE Transactions on Visualization and Computer Graphics},
  23(1):291--300, 2016.

\bibitem{yu}
H.~Yu.
\newblock Spatio-temporal gis design for exploring interactions of human
  activities.
\newblock {\em Cartography and Geographic Information Science}, 33(1):3--19,
  2006.

\bibitem{yuan2014visualization}
X.~Yuan, Z.~Wang, Z.~Liu, C.~Guo, H.~Ai, and D.~Ren.
\newblock Visualization of social media flows with interactively identified key
  players.
\newblock In {\em IEEE Conference on Visual Analytics Science and Technology},
  pp. 291--292, 2014.

\bibitem{zgraggen2015s}
E.~Zgraggen, S.~M. Drucker, D.~Fisher, and R.~DeLine.
\newblock (s|qu) eries: Visual regular expressions for querying and exploring
  event sequences.
\newblock 2015.

\bibitem{zhang2020cause}
W.~Zhang, T.~K. Panum, S.~Jha, P.~Chalasani, and D.~Page.
\newblock Cause: Learning granger causality from event sequences using
  attribution methods.
\newblock {\em arXiv preprint arXiv:2002.07906}, 2020.

\bibitem{zhang2018idmvis}
Y.~Zhang, K.~Chanana, and C.~Dunne.
\newblock Idmvis: Temporal event sequence visualization for type 1 diabetes
  treatment decision support.
\newblock {\em IEEE Transactions on Visualization and Computer Graphics},
  25(1):512--522, 2018.

\bibitem{zhang2015iterative}
Z.~Zhang, D.~Gotz, and A.~Perer.
\newblock Iterative cohort analysis and exploration.
\newblock {\em Information Visualization}, 14(4):289--307, 2015.

\bibitem{zhang2014visual}
Z.~Zhang, K.~T. McDonnell, E.~Zadok, and K.~Mueller.
\newblock Visual correlation analysis of numerical and categorical data on the
  correlation map.
\newblock {\em IEEE Transactions on Visualization and Computer Graphics},
  21(2):289--303, 2014.

\bibitem{zhao2014fluxflow}
J.~Zhao, N.~Cao, Z.~Wen, Y.~Song, Y.-R. Lin, and C.~Collins.
\newblock \# fluxflow: Visual analysis of anomalous information spreading on
  social media.
\newblock {\em IEEE Transactions on Visualization and Computer Graphics},
  20(12):1773--1782, 2014.

\bibitem{zhao2015matrixwave}
J.~Zhao, Z.~Liu, M.~Dontcheva, A.~Hertzmann, and A.~Wilson.
\newblock Matrixwave: Visual comparison of event sequence data.
\newblock In {\em Proceedings of the CHI Conference on Human Factors in
  Computing Systems}, pp. 259--268, 2015.

\bibitem{zhou2019survey}
F.~Zhou, X.~Lin, C.~Liu, Y.~Zhao, P.~Xu, L.~Ren, T.~Xue, and L.~Ren.
\newblock A survey of visualization for smart manufacturing.
\newblock {\em Journal of Visualization}, 22(2):419--435, 2019.

\bibitem{zhou2018visually}
F.~Zhou, X.~Lin, X.~Luo, Y.~Zhao, Y.~Chen, N.~Chen, and W.~Gui.
\newblock Visually enhanced situation awareness for complex manufacturing
  facility monitoring in smart factories.
\newblock {\em Journal of Visual Languages \& Computing}, 44:58--69, 2018.

\end{thebibliography}

\end{spacing}
\end{document}